\documentclass[letterpaper,12pt]{article}

\usepackage[linkcolor=blue,urlcolor=blue,citecolor=blue, colorlinks,linktocpage,bookmarks,breaklinks,pdfstartview=FitH]{hyperref}
\usepackage{natbib} 
\usepackage{epsfig}
\usepackage{graphicx}
\usepackage{fancyhdr}
\usepackage[all]{hypcap}
\usepackage{latexsym}
\usepackage{amsmath}
\usepackage{verbatim}
\usepackage{setspace}
\usepackage{subfig}
\usepackage{titlesec}
\usepackage{ulem}
\usepackage{paralist}

\allowdisplaybreaks
\makeatletter
\g@addto@macro\normalsize{
\setlength\abovedisplayskip{5pt}
\setlength\belowdisplayskip{5pt}
}
\makeatother

\def\mb#1{\mbox{\boldmath $#1$}}
\def\bb {\mb b}
\def\by {\mb y}
\def\btheta {\mb \theta}
\def\bmu {\mb \mu}
\def\bsgm {\mb \sigma}

\def \obs {^{\mbox{\tiny obs}}}
\def \noi {_{-i}}
\def \wi {_{1:n}}

\def \CVIC {\mbox{CVIC}}

\def \cvpostnoM {_{\mbox{\scriptsize post(-i)}}}
\def \QcvpostnoM {P\cvpostnoM(\btheta, \bb\wi | \by\obs\noi)}
\def \cvpostM {_{\mbox{\scriptsize post(-i), M}}}
\def \QcvpostM {P\cvpostM(\btheta, \bb\noi | \by\obs\noi)}
\def \postfnoM {_{\mbox{\scriptsize post}}}
\def \QpostfnoM {P\postfnoM (\btheta, \bb\wi | \by\obs\wi)}
\def \PpostfnoM {P (\btheta, \bb\wi | \by\obs\wi)}
\def \postfM {_{\mbox{\scriptsize post, M}}}
\def \QpostfM{P_{\mbox{\scriptsize post, M}} (\btheta, \bb\noi | \by\obs\noi)}
\def \nIS {^{\mbox{\scriptsize nIS}}}
\def \iIS {^{\mbox{\scriptsize iIS}}}
\def \mcs {^{(s)}}
\def \waic {^{\mbox{\scriptsize WAIC}} }
\def \nwaic {^{\mbox{\scriptsize nWAIC}} }
\def \iwaic {^{\mbox{\scriptsize iWAIC}}}

\begin{document}

\bigskip 

\begin{center}

\doublespacing

{\Large \bf Approximating Cross-validatory Predictive Evaluation in Bayesian Latent Variables Models with Integrated IS and WAIC}

\onehalfspacing

{Longhai Li\footnotemark[1]\footnotemark[3],  Shi Qiu\footnotemark[1], Bei Zhang\footnotemark[1], and Cindy X. Feng\footnotemark[2]}

%\textit{Manuscript,} 
15 January 2015

\end{center}

% !TEX root = iis.tex

\onehalfspacing \noindent \textbf{Abstract:} Looking at predictive accuracy is a traditional method for comparing  models. A natural method for approximating out-of-sample predictive accuracy is leave-one-out cross-validation (LOOCV) --- we alternately hold out each case from a full data set and then train a Bayesian model using Markov chain Monte Carlo (MCMC) without the held-out; at last we evaluate the posterior predictive distribution of all cases with their actual observations. However, actual LOOCV is time-consuming. This paper introduces two methods, namely iIS and iWAIC, for approximating LOOCV with only Markov chain samples simulated from a posterior based on a \textit{full} data set. iIS and iWAIC aim at improving the approximations given by importance sampling (IS) and WAIC in Bayesian models with possibly correlated latent variables. In iIS and iWAIC, we first integrate the predictive density over the distribution of the latent variables associated with the held-out without reference to its observation, then apply IS and WAIC approximations to the integrated predictive density. We compare iIS and iWAIC with other approximation methods in three kinds of models: finite mixture models, models with correlated spatial effects, and a random effect logistic regression model. Our empirical results show that iIS and iWAIC give substantially better approximates than non-integrated IS and WAIC and other methods.   

\noindent \textbf{Key phrases:} MCMC, cross-validation, posterior predictive check, predictive model assessment, DIC, WAIC, Bayesian latent variable models

\singlespacing

\footnotetext[1]{  
Department of Mathematics and Statistics, University of Saskatchewan, 106 Wiggins Rd, Saskatoon, SK, S7N5E6, Canada. E-mails: \texttt{longhai@math.usask.ca}, \texttt{shq471@mail.usask.ca}, \texttt{bez733@mail.usask.ca}. }

\footnotetext[2]{
School of Public Health and Western College of Veterinary Medicine, University of Saskatchewan, Health Sciences Building, 107 Wiggins Road, Saskatoon, SK, S7N 5E5 Canada. Email: \texttt{cindy.feng@usask.ca}. }

\footnotetext[3]{Correspondance author.}

\newpage

% !TEX root =  iis.tex

\doublespacing
% control space around section
\titlespacing*{\section}{0pt}{-1pt}{0pt}
\titlespacing*{\subsection}{0pt}{-1pt}{0pt}

\section{Introduction} \label{sec:intro} 

Evaluating goodness of fit of models to a data set is of fundamental importance in statistics. The goodness-of-fit evaluation is necessary for many tasks, such as, comparing competing models (which may be non-nested), testing hypotheses,  and detecting outliers in a data set to a model. To date, evaluating model goodness-of-fit remains a daunting problem for Bayesian statisticians. There have been a wide range of methods for this problem, in addition to the classic significance test for parameters used to link a family of nested models.  In particular, Bayes factor \citep{kass1995bayes} based on marginal likelihood is widely used for comparing multiple Bayesian models. However, it is notorious that marginal likelihood can be arbitrarily small if the prior is sufficiently diffuse --- a problem called Jeffrey-Lindley paradox \citep{lindley1957statistical, robert2013jeffreys-lindleys}, therefore Bayes factor cannot be used in models with uninformative or improper priors. Much research has been done to remedy this problem with various methods; to name a few, the fractional Bayes factor \citep{ohagan1995fractional, ohagan1997properties}, the intrinsic Bayes factor \citep{berger1996intrinsic}, and the methods treating model selection as a decision problem and using continuous loss functions, see \citet{bernardo2002bayesian, li2012bayesian, li2014new}, and the references therein. In addition, computing marginal likelihood is tremendously difficult for complex models, see discussion in \citet{chib1995marginal, raftery2006estimating},  and the references therein.   Another traditional approach, often called predictive model assessment, is to look at accuracy of competing models in predicting out-of-sample observations, which is free of Jeffrey-Lindley paradox.  An extensive review of predictive model assessment methods is provided by  \cite{vehtari2012survey}. 

Cross-validation (CV) is a natural way to approximate out-of-sample predictive performance of a model. Throughout this paper, we will discuss only leave-one-out cross-validation (LOOCV); hence in what follows, CV means LOOCV.  In CV, we hold out a unit from a full data set, fit/train a model using Markov chain Monte Carlo (MCMC) without the holdout, and then find a predictive distribution of what would be observed from the holdout. We repeat this procedure with each observation as a holdout.  We can then compare the CV predictive distributions with the actual observations  in terms of a chosen loss function.  A widely used loss function is negative twice log predictive density of the actual observation. Predictive evaluations based on this loss are often called  \textit{information criteria} (IC) for historical reason \citep{gelman2014understanding}.  CV predictive evaluation can also be used to check whether the actual observation is an outlier by looking at tail probability of the predictive distribution \citep{SIM:SIM1403, marshall2007identifying}.    Actual Bayesian CV is time-consuming for complex models because it requires an MCMC simulation for each unit as a held-out test case. Alternative methods have been proposed to approximate out-of-sample or CV predictive evaluations only with MCMC samples drawn from the posterior based on the full data set. These methods aim at correcting for optimistic bias in training (also called within-sample) predictive evaluation. \citet{gelfand92bs} introduce importance sampling (IS) method that weights MCMC samples using inverse training predictive density for each unit. IS is widely applicable to many loss functions. This method has been innovatively applied to many problems, such as in off-policy reinforcement learning problems \citep{hachiya2008adaptive} and in ``inverse problems'' \citep{bhattacharya2007importance}. However, many applications show that IS approximation has large bias and variance \citep{peruggia1997variability, vehtari2001bayesian, vehtari2002cv, epifani2008case}. 

There are also many  other methods that focus on estimating out-of-sample information criterion by adjusting a version of training predictive information criterion with a correction for optimistic bias \citep{spiegelhalter2002bayesian, Ando01062007, plummer2008penalized, gelman2014understanding}. In the recent years, the deviance information criterion (DIC) of  \cite{spiegelhalter2002bayesian} may be the most popular choice  in Bayesian applications, which is readily available in WinBUGS. However, a number of difficulties have been noted with DIC (and its variants), particularly in Bayesian models in which latent variables and model parameters are non-identifiable from data --- a typical example is mixture models; see \cite{celeux2006deviance}, and \cite{plummer2008penalized}, and many of the discussions following the paper by \cite{spiegelhalter2002bayesian}. Some authors have pointed out connections and discrepancies of DIC with out-of-sample information criterion [see  \cite{plummer2008penalized, watanabe2010asymptotic, gelman2014understanding}]. However,  nowadays we often need to compare models with latent variables.  DIC is typically implemented by treating latent variables as unknown parameters otherwise DIC will be too hard to implement;  however, this treatment is lack of theoretical justification; see a detailed discussion in \citet{li2012robust}.  Recently, a newer criterion called WAIC (widely applicable information criterion) was proposed by \citet{watanabe2009algebraic, watanabe2010equationsb, watanabe2010equationsa}, which has been evaluated in several simple models by \citet{gelman2014understanding}. WAIC operates on predictive probability density of observed variables rather than on model parameters,  hence, it can be applied in singular statistical models (ie, models with non-identifiable parameterization). \citet{watanabe2010asymptotic} has proved that WAIC is equivalent to CV information criterion asymptotically as random variables of training data, and that on average of both training and evaluation (future) data, both WAIC and CV information criterion are asymptotically equivalent to out-of-sample information criterion using his singular statistical learning theory \citep{watanabe2009algebraic}. However, WAIC is only justified for problems where observed data are independently distributed with a population distribution.

In this article, we introduce two predictive evaluation methods based on IS and WAIC  for use in Bayesian models with unit-specific and possibly correlated latent variables.  IS and WAIC can be simply applied to the (non-integrated) predictive density of observed variables, which is conditional not only on the model parameters, but also latent variables associated with a validation unit that is supposed to be left out in CV.  However, the actual observations on the validation unit used in full data posterior often bring more bias into the latent variables associated with the validation unit (perhaps more than into the model parameters) than IS or WAIC correction alone can eliminate.  To eliminate the bias in the latent variables associated with the validation unit,  one remedy is to temporarily discard the latent variables in full data posterior sample, and integrate the non-integrated predictive density with respect to the conditional distribution of the latent variables associated with the validation unit that is conditional on only the model parameters \textit{but not the actual observations}, which will lead to an \textit{integrated} predictive density.  Using the same way we obtain integrated evaluation function. We then apply IS and WAIC formulae to the integrated predictive density and evaluation functions, which results in two predictive evaluation methods --- Integrated Importance Sampling (iIS) and Integrated WAIC (iWAIC).  The required integrals can be obtained analytically in some models, otherwise, can often be easily approximated using Monte Carlo methods or other numerical methods.  

\citet{vehtari2001bayesian, vanhatalo2012bayesian, vanhatalo2013gpstuff:} have used iIS for computing information criterion, a special but very important case of predictive evaluation, in Gaussian process latent variable models in their matlab toolbox \texttt{GPstuff}. For computing information criterion,  one uses only the integrated predictive density (see equation \eqref{eqn:pdhmmM}), for which \texttt{GPstuff} used analytical method for Gaussian likelihood, and numerical approximation for non-Gaussian likelihood; this is documented by the manual for \texttt{GPstuff} but their technical report \citep{vanhatalo2012bayesian} did not discuss the details of iIS.   Our article gives iIS a detailed discussion. In addition, we provide a formula for iIS that is applicable to general evaluation function;  in particular,  our formula can be used also for computing CV posterior p-value. We have also proved the equivalence of iIS and actual CV. The main contribution of this paper is to use illustrative examples to demonstrate the necessity and benefit of integrating away the latent variables associated with the validation unit.  For computing CV posterior p-value,  iIS is also related to another method called ghosting method,  which was proposed by \citet{marshall2007identifying}, and also discussed by \citet{held2010posterior}. Ghosting method discards latent variables associated with the validation unit and re-generates them from the distribution without reference to the actual observations of the validation unit using Monte Carlo method to compute a tail probability (evaluation function), but ghosting method does not use importance re-weighting to correct for the bias in model parameters; hence, ghosting method can be deemed as a partial implementation of iIS.  

The remaining of this article will be organized as follows. In Section 1, we describe a class of Bayesian models with unit-specific models that iIS and iWAIC can be applied to.  In Section 2, we describe how to perform actual cross-validation evaluation,  and give relevant posterior distributions.  We will then describe iIS and iWAIC in general terms in Sections \ref{sec:iIS} and \ref{sec:iWAIC}, respectively.  In Section \ref{sec:examples},  we compare iIS and iWAIC to other approximation methods in three simple examples --- a mixture modelling problem, a problem using random effect logistic models, and a disease mapping problem that uses spatially correlated random effects. Our empirical results show that iIS and iWAIC provide significantly closer approximates to actual CV evaluation results than ordinary IS and WAIC, as well as other methods. 
 The article will be concluded in Section \ref{sec:conc}.  In Appendices, we give a sketch of the working procedures of iIS and iWAIC.

\section{Bayesian Models with Unit-specific Latent Variables} \label{sec:bmlv}

\begin{figure}[tp]

\caption{Graphical representation of Bayesian latent variables models. The double arrows in the box for $\bb_{1:n}$ mean possible dependency between $\bb_{1:n}$. Note that the covariate $\mb x_i$ will be omitted in the conditions of densities for $\bb_i$ and $\by_i$ throughout this paper for simplicity. } 

\label{fig:blvm}

\centering
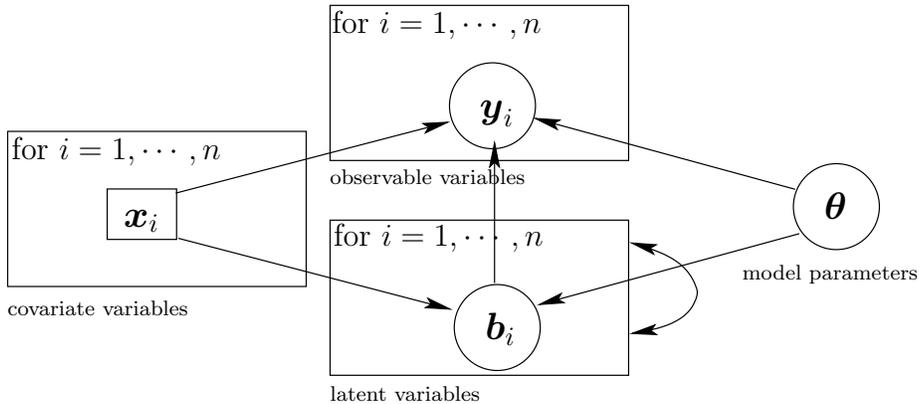
\end{figure}

The new predictive evaluation methods that we will describe is for use in Bayesian models with unit-specific latent variables. Throughout this paper, we use bold-faced letters to denote vectors and matrices. Suppose we have $n$ observations $\by_1\obs, \cdots, \by_n\obs$ on $n$ observation units (aka cases, such as persons, locations, time points, or a combination of them). We model them as a realization of random variables $\by_1,\cdots,\by_n$.  In many problems, we introduce a latent variable (often random vector, sometimes called random effects, missing data) $\bb_i$ for each unit $i$ from which $\by_i\obs$ is observed, then we will model $\by_i$ and $\bb_i$ with certain statistical distributions parametrized by $\btheta$.  Conditional on $\bb_i$ and $\btheta$ (often also on a covariate variable $\mb x_i$ that will be omitted in following equations for simplicity), we assume that $\by_1,\cdots, \by_n$ are statistically independent, with probability density $P(\by_i|\bb_i, \btheta)$, which we will call \textbf{non-integrated predictive density} in this article. If we assume independence between $\bb_1,\cdots, \bb_n$ given $\btheta$, then the marginalized distributions of random variables $\by_1,\cdots, \by_n$ given $\btheta$ are also independent for each $i$, for example in mixture models. For modelling spatial and time series data, we often assume that the latent variables $\bb_1,\cdots, \bb_n$ are dependent for modelling correlations between locations or time points (see an example in Section \ref{sec:spatial}).  In the following general discussion, we will assume that $\bb_1,\ldots, \bb_n$ are correlated. Figure \ref{fig:blvm} gives a graphical representation of the models described here. 

Throughout this paper, we will use notation $\mb a_{1:n}$ to denote the collection of all $\mb a_j$: $\{\mb a_j | j = 1,\ldots,n\}$, and use $\mb a_{-i}$ to denote the collection of all $\mb a_j$ except $\mb a_i$: $\{\mb a_j | j = 1, \ldots,n, j \not=i\}$. 

Suppose conditional on $\btheta$, we have specified a density for $\by_i$ given $\bb_i$: $P(\by_i|\bb_i, \btheta)$, a joint prior density for latent variables $\bb_{1:n}$: $P(\bb_{1:n}|\btheta)$, and a prior density for $\btheta$: $P(\btheta)$. The posterior of $(\bb_{1:n}, \btheta)$ given observations $\by_{1:n}\obs$ is proportional to the joint density of $\by_{1:n}\obs$, $\bb_{1:n}$, and $\btheta$:
\begin{equation}
\QpostfnoM
=   \prod_{j=1}^n P(\by_j\obs|\bb_j,\btheta) P(\bb_{1:n}|\btheta) P(\btheta)/C_{1}, \label{eqn:jointfull}
\end{equation}
where $C_{1}$ is the normalizing constant involving only with $\by\wi\obs$.

\section{Actual Cross-validatory Predictive Evaluation}

To do cross-validation, for each $i = 1,\ldots, n$, we omit observation $\by_i\obs$, and then draw MCMC samples from \textbf{CV posterior distribution} of model parameter and latent variables $P(\btheta, \bb_{1:n}|\by_{-i}\obs)$:
\begin{equation}
\QcvpostnoM  
=   \prod_{j \not = i} P(\by_j\obs|\bb_j,\btheta) P(\bb_{1:n}|\btheta) P(\btheta)\,/\,C_{2}, \label{eqn:jointCV}
\end{equation}
where $C_{2}$ is the normalizing constant involving only with $\by\noi\obs$.  Note that, in equation \eqref{eqn:jointCV}, we assume that the possible structures information (e.g. spatial relationships between $n$ locations) among $\bb_{1:n}$ are not lost, with only the value of $\by_{i}\obs$ omitted. After we draw MCMC samples of $(\btheta, \bb\wi)$ from \eqref{eqn:jointCV}, and then drop $\bb_i$, we obtain MCMC sample of $(\btheta, \bb_{-i})$ from the marginalized CV posterior $P(\btheta, \bb_{-i} | \by_{-i}\obs)$:
\begin{equation}
 \QcvpostM  
=   \prod_{j \not = i} P(\by_j\obs|\bb_j,\btheta) P(\bb\noi|\btheta) P(\btheta)\,/\,C_{2}, \label{eqn:mjointCV}
\end{equation}
where $P(\bb\noi|\btheta)$ is the marginalized prior density for $\bb\noi$ induced from the specified joint prior for $\bb\wi$,  i.e., $P(\bb\noi|\btheta) = \int P(\bb\wi|\btheta)d\bb_{i}$. Using conditional prior $P(\bb_i|\bb\noi,\btheta) = P(\bb\wi| \btheta)/P(\bb\noi|\btheta)$, we can write 
\begin{equation}
\label{eqn:MnoM}
\QcvpostnoM =  \QcvpostM P(\bb_i|\bb\noi, \btheta).
\end{equation}

From the above expression,  we see that sampling from $P\cvpostnoM$ is equivalent to sampling from $P\cvpostM$ and then generating $\bb_i$ from the conditional prior $P(\bb_i|\bb\noi,\btheta)$. Therefore, this way to perform cross-validation makes use of the assumed structure in $\bb_{1:n}$ (such as neighbouring relationships between spatial units, see the example presented in Section \ref{sec:spatial}) through $P(\bb_i|\bb\noi,\btheta)$, in predicting $\by_i$ given $\by\noi\obs$.  This treatment indeed regards the structure information in $\bb_{1:n}$ as fixed covariate and being known. We feel that this treatment is reasonable because we are interested in comparing competing models for the conditional distribution of $\by_{1:n}$ given the structure between the $n$ units, rather than the distribution of the structure itself.  This is similar to how the cross-validation is done in linear models, for which we assume that the values of the covariates (explanatory variables) of the test case are known when we make prediction of the response of the test case. 

The purpose of performing CV is to evaluate certain compatibility (or discrepancy) between the posterior $P(\by_i|\by\noi\obs)$ and the actual observation $\by_{i}\obs$. We will specify an evaluation function $a (\by\obs_i,\btheta,\bb_i)$ that measures certain goodness-of-fit (or discrepancy) of  the distribution $P(\by_i | \btheta, \bb_i)$ to the actual observation $\by\obs_i$. \textbf{CV posterior predictive evaluation} is defined as the expectation of the $a (\by\wi\obs, . , .)$ with respect to $P\cvpostnoM$: 
\begin{equation} 
 \label{eqn:cvevalnoM}
 E\cvpostnoM (a (\by_i\obs, \btheta,\bb_i)) 
 = \int a (\by_i\obs, \btheta,\bb_i ) P\cvpostnoM(\btheta, \bb_{1:n}|\by\noi\obs)d\btheta d\bb\wi.
\end{equation}
The expectation in \eqref{eqn:cvevalnoM} can be approximated by averaging $a(\by\obs_i,\cdot,\cdot)$ over MCMC samples of $(\btheta, \bb_i)$ drawn from $P\cvpostnoM$. 

The first example of $a$ is the value of predictive density function $P(\by_{i}|\bb_i, \btheta)$ at the actual observation $\by_{i}\obs$: 
\begin{equation}
a (\by_i\obs, \btheta,\bb_i) = P(\by_i\obs|\btheta, \bb_i). \label{eqn:icpw}
\end{equation}
The expectation of \eqref{eqn:icpw} with respect to $P\cvpostnoM$ is \textbf{CV posterior predictive density} $P(\by_i\obs|\by\noi\obs)$.  \textbf{CV information criterion} (CVIC) is defined as the sum of minus twice of CV posterior predictive densities over all validation units:
\begin{equation}
\CVIC = -2\sum_{i=1}^n\log (P(\by_i\obs|\by\noi\obs)). \label{eqn:cvic}                                                                                                                                                                                                                                                                                                                                                                                                                                                                                                                                                                                                                      \end{equation}
A smaller value of CVIC indicates a better fit of a Bayesian model to a real data set. The second is to set $a$ in \eqref{eqn:cvevalnoM} as the p-value given model parameter and latent variable for unit $i$ \citep{SIM:SIM1403, marshall2007identifying}:
\begin{equation}
a (\by_i\obs, \btheta,\bb_i) = Pr(\by_i > \by_i\obs | \btheta, \bb_i) + 0.5  Pr(\by_i = \by_i\obs | \btheta, \bb_i), 
\label{eqn:ppvpw}
\end{equation}
where $Pr$ means probability of a set, as we have used $P$ as density; also $\by_i$ should be a scalar for such situations. The expectation of \eqref{eqn:ppvpw} with respect to $P\cvpostnoM$ gives \textbf{CV posterior p-value}:
\begin{equation}
 \mbox{CV posterior p-value}\ (\by\obs_i|\by\noi\obs) = Pr(\by_i > \by_i\obs |\by\noi\obs) + 0.5 Pr(\by_i = \by_i\obs |\by\noi\obs),\label{eqn:ppv}
\end{equation}
which is a tail probability of CV posterior predictive distribution with density $P(\by_{i}|\by\noi\obs)$.  The purpose of computing CV posterior p-value is to check the discrepancy of the observation $\by_i\obs$ to the CV posterior predictive distribution of $\by_i$ that is conditional on other observations $\by\noi\obs$. Both very large and small values of posterior p-value indicate that $\by_i\obs$ may be an outlier (unusually small or large) compared to other observations.

Actual CV requires $n$ of Markov chain simulations (each may use multiple parallel chains), one for each validation unit. This is very time consuming, especially when the model is complex and $n$ is fairly large. Therefore, we are interested in approximating the expectations in \eqref{eqn:cvevalnoM} for all validation units $i=1,\ldots, n$ with samples of $(\btheta, \bb\wi)$ obtained with a single MCMC simulation based on the full data set; that is, with samples drawn from $\QpostfnoM$, called \textbf{full data posterior} for short hereafter. However, we cannot simply treat samples from the full data posterior as CV posteriors, because the inclusion of $\by_{i}\obs$ has introduced optimistic bias in validating $\by_{i}\obs$. The optimistic bias means that the ``posterior predictive distribution'' of $\by_{i}$ formed by averaging $P(\by_{i}|\bb_{i}, \btheta)$ with respect to $\QpostfnoM$ fits $\by_{i}\obs$ better than the actual CV posterior predictive distribution of $\by_{i}$ that averages $P(\by_{i}|\bb_{i}, \btheta)$ with respect to $\QcvpostnoM$. Therefore, we need to correct for the optimistic bias with a certain method to obtain an unbiased approximate/estimate of actual CV posterior predictive evaluation. We will introduce two new approximating methods in Section \ref{sec:iIS} and \ref{sec:iWAIC}, respectively.

\section{Importance Sampling (IS) Approximation}\label{sec:iIS}

\subsection{Non-integrated Importance Sampling} \label{sec:is}

Importance weighting \citep{gelfand92bs} is a natural choice for approximating CV prediction evaluation based on the posterior given the full data set. For general and detailed discussion of importance sampling techniques, one can refer to \citet{geweke1989bayesian,  neal:1993, gelman1998simulating, liu:mcbook}. If our samples are from $\QpostfnoM$, but we are interested in estimating the mean of $a$ with respect to $\QcvpostnoM$ as in \eqref{eqn:cvevalnoM},  importance weighting method is based on the following equality for CV expected evaluation: 
\begin{equation}
 \label{eqn:isenoM} 
 E\cvpostnoM (a(\by_i\obs,\btheta,\bb_i)) =
 \dfrac{E\postfnoM\big[ a(\by_i\obs,\btheta,\bb_i) W_{i}\nIS(\btheta,\bb\wi)\big]}
 {E\postfnoM \big[W_{i}\nIS(\btheta,\bb\wi)\big]},
\end{equation}
where $E\postfnoM[\ ]$ is expectation with respect to $\QpostfnoM$, and 
\begin{equation}
 \label{eqn:imwnoM} W_{i}\nIS(\btheta,\bb\wi)  = \frac {\QcvpostnoM} {\QpostfnoM} \times \frac{C_2}{C_1} = \frac{1}{P(\by_i\obs|\btheta, \bb_i)}. 
\end{equation}
Note that, we can multiply any constant to the above important weight since they will be canceled in the fraction of \eqref{eqn:isenoM}; also we use subscript $\nIS$ denote application of importance sampling (shortened by \textbf{nIS}) to the \textbf{non-integrated predictive density}, in contrast to iIS to be given in next section.  In words, important sampling estimates the expected evaluation by finding Monte Carlo estimates of the two means in the fraction of \eqref{eqn:isenoM} with only MCMC samples from $\QpostfnoM$. We can apply equation \eqref{eqn:isenoM} to estimate means of any evaluation function $a$ with respect to the CV posterior distribution of $(\btheta, \bb_i)$. 

Particularly, in computing CVIC, the evaluation function $a(\by_i\obs,\btheta,\bb_i) = P(\by_i\obs|\btheta, \bb_i)$, which is the same as $1/W_{i}\nIS(\btheta,\bb\wi)$ in equation \eqref{eqn:imwnoM}. Therefore, the numerator of \eqref{eqn:isenoM} is just 1 when applied to compute CVIC. Therefore, the CV posterior predictive density $P(\by_i\obs|\by\noi\obs)$ is equal to harmonic mean of the non-integrated predictive density $P(\by_i\obs|\btheta, \bb_i)$ with respect to $\PpostfnoM$: 
\begin{equation}
 \label{eqn:hmm-nis}
 P(\by_i\obs|\by\noi\obs) = \dfrac{1}{E\postfnoM \big[1/P(\by_i\obs|\btheta,\bb_i)\big]}.
\end{equation}
Based on the equality \eqref{eqn:hmm-nis}, \textbf{nIS} estimates the CV posterior predictive density by:  
\begin{equation}
 \label{eqn:hmmicnoM}
 \hat P \nIS (\by_i\obs|\by\noi\obs) = \dfrac{1}{\hat E\postfnoM \big[1/P(\by_i\obs|\btheta,\bb_i)\big]}.
\end{equation}
The corresponding nIS estimate of CVIC using \eqref{eqn:hmmicnoM} is $-2\sum_{i=1}^n\log (\hat{P}\nIS(\by_i\obs|\by\noi\obs)$.  Note that,  if there are not latent variables used for a model,  there will be no $\bb_i$  in \eqref{eqn:hmm-nis} and  \eqref{eqn:hmmicnoM}.  

\subsection{Integrated Importance Sampling} \label{sec:iIS2}

In theory, the nIS estimate \eqref{eqn:isenoM} is valid for almost all Bayesian models with latent variables as long as the integral itself exists and the supports of $\QcvpostnoM$ and $\QpostfnoM$ are the same. However, in simulating MCMC from $\QpostfnoM$, the latent variable $\bb_i$ is largely confined to regions that fit well the observation $\by_i\obs$. Therefore, the distribution of $\bb_i$ marginalized from $\QpostfnoM$ may be highly biased to regions that fit well the observation $\by_i\obs$, compared to the distribution of $\bb_i$ marginalized from  $\QcvpostnoM$, which can cover a much larger area. Therefore, although the supports of  $\QcvpostnoM$ and $\QpostfnoM$ are the same in theory, the effective support of $\QpostfnoM$ may be much smaller than that of $\QcvpostnoM$. We will illustrate this in the mixture model example with Figure \ref{fig:nint}.  This results in the inaccuracy of nIS.  

To improve nIS, we can re-generate $\bb_i$ from $P(\bb_i|\bb\noi, \btheta)$, with the observation $\by\obs_i$ removed, as the actual cross-validation simulation does; see equation (\ref{eqn:MnoM}).  The formal formulation of such re-generation procedure is given as follows. First we note that using equation \eqref{eqn:MnoM}, we can rewrite the expectation in \eqref{eqn:cvevalnoM} as 
\begin{eqnarray}
 \lefteqn{E\cvpostnoM (a (\by_i\obs, \btheta,\bb_i))  = E\cvpostM (A (\by_i\obs, \btheta, \bb\noi))} \\
 &=&\int \int  A (\by_i\obs, \btheta, \bb\noi) P(\btheta, \bb\noi|\by\noi\obs) d\btheta d\bb\noi, 
 \label{eqn:cvevalM} 
\end{eqnarray}
where,
\begin{equation}
\label{eqn:intA}
A (\by_i\obs, \btheta, \bb\noi) = \int a (\by_i\obs, \btheta,\bb_i ) P(\bb_i |\bb\noi, \btheta)d\bb_i. 
\end{equation}
We will call \eqref{eqn:intA} as an \textbf{integrated evaluation function}. 

We will also discard $\bb_i$ temporarily for validation unit $i$ in MCMC samples from the full data posterior $\QpostfnoM$. The marginalized full data posterior of $(\btheta,\bb\noi)$ is
\begin{equation}
 \QpostfM =  \prod_{j\not=i} P(\by_j\obs|\bb_j,\btheta) P(\bb\noi|\btheta) P(\btheta) \! \times \! \int \!P(\by_i\obs|\bb_i,\btheta)P(\bb_i|\bb\noi,\btheta)d \bb_i / C_{1}.  
\label{eqn:postfullM}
\end{equation}
We will call the second factor in \eqref{eqn:postfullM} \textbf{integrated predictive density}, because it integrates away $\bb_i$ without reference to $\by_i\obs$. For ease in reference, it is explicitly given below:
\begin{equation}
 \label{eqn:inpd}
 P(\by_i\obs|\btheta, \bb\noi) = \int P(\by_i\obs|\bb_i,\btheta)P(\bb_i|\bb\noi,\btheta)d \bb_i.
\end{equation}
Using the standard importance weighting method,  we will estimate \eqref{eqn:cvevalM} by 
\begin{equation}
 \label{eqn:iseM} \
 E\cvpostM (A(\by_i\obs,\btheta,\bb\noi)) =
 \dfrac{E\postfM\big[ A(\by_i\obs,\btheta,\bb\noi)\ W_{i}\iIS(\btheta,\bb\noi)\big]}
 {E\postfM \big[W_{i}\iIS(\btheta,\bb\noi)\big]},
\end{equation}
where $W_{i}\iIS$ is the integrated importance weight:
\begin{equation}
 \label{eqn:iswM} W_{i}\iIS (\btheta,\bb\noi) = 
 \dfrac{\QcvpostM}{\QpostfM} \times \dfrac{C_{2}}{C_{1}} = \dfrac{1}{P(\by_i\obs|\btheta, \bb\noi)}.
\end{equation}

In particular, for estimating CVIC, $A \times W_{i}\iIS = 1$. Therefore, the iIS estimate of the CV posterior predictive density based on equality \eqref{eqn:iseM} is given by:
\begin{equation}
\label{eqn:pdhmmM}
 \hat P \iIS (\by_i\obs|\by\noi\obs) = \dfrac{1}{\hat E\postfM\big[1/P(\by_i\obs|\btheta, \bb\noi)\big]}.
\end{equation}
Accordingly, iIS estimate of CVIC using \eqref{eqn:pdhmmM} is $-2\sum_{i=1}^n \log (\hat{P}\iIS(\by_i\obs|\by\noi\obs))$. The difference from nIS estimate \eqref{eqn:hmmicnoM} is only the replacement of non-integrated predictive density $P(\by\obs_i|\btheta,\bb_i)$ by integrated predictive density $P(\by\obs_i|\btheta, \bb\noi)$. Note that we can also write the expectation $E\postfM(\ )$ in equations \eqref{eqn:iseM} and (\ref{eqn:pdhmmM}) as $E\postfnoM(\ )$, because we still find Monte Carlo estimates with samples of $(\btheta, \bb\wi)$ from $\QpostfnoM$, but without using $\bb_{i}$.  

The integration over $\bb_i$ in equations \eqref{eqn:intA} and \eqref{eqn:inpd} is the essential difference of iIS to nIS. For using iIS, we need to find their values. In some problems, they can be approximated with finite summation, or calculated analytically. Otherwise, we will re-generate $\bb_i$ given $(\bb\noi, \btheta)$ with no reference to $\by_i\obs$, which is often easy. Note that this re-generation needs to be done for each $i=1,\ldots,n$. Sometimes, much computation can be shared by these $n$ re-generating processes since they are all conditional on $\btheta$; see the example in Section \ref{sec:spatial}.

\section{WAIC Approximations} \label{sec:iWAIC}

In this section, we describe a generalized WAIC method, iWAIC, for approximating CV predictive density in Bayesian models with correlated latent variables. 

% \subsection{WAIC for Models without Latent Variables}

We will first describe WAIC for models with no latent variables (or models after we integrate away latent variables that are independent for units given parameters). In such models, observed variables $\by_1,\ldots,\by_n$ are independently distributed with a probability distribution $P(\by|\btheta)$ conditional on model parameters $\btheta$. After we obtain MCMC samples for $\btheta$ given observations $\by_1\obs,\ldots,\by_n\obs$, a version of WAIC \citep{watanabe2009algebraic, watanabe2010equationsb, watanabe2010equationsa} is given by:
\begin{equation}
 \label{eqn:waicind}
 \mbox{WAIC} = -2\sum_{i=1}^n\big[ \log (E\postfnoM(P(\by_i\obs|\btheta))) - V\postfnoM(\log (P(\by_i\obs|\btheta))) \big],
\end{equation}
where $E\postfnoM$ and $V\postfnoM$ stand for mean and variance over $\btheta$ with respect to $P(\btheta|\by_1\obs,\ldots, \by_n\obs)$. By comparing the forms of WAIC and CVIC \eqref{eqn:cvic}, we can think of that in WAIC,  the CV posterior predictive density is approximated by:
\begin{equation}
 \label{eqn:pdwaic}
 \hat{P}\waic (\by_i\obs|\by\noi\obs) = \exp\big\{\log (E\postfnoM(P(\by_i\obs|\btheta))) - V\postfnoM(\log (P(\by_i\obs|\btheta)))\big\}. 
\end{equation}
In words, WAIC corrects for the bias in mean of training predictive density of $\by_i\obs$ by dividing exponential of variance of log predictive density of $\by_i\obs$ with respect to the posterior of $\btheta$ given the full data set. \citet{watanabe2010asymptotic} has proven that WAIC is asymptotically equivalent to CVIC when observed variables are independently distributed conditional on $\btheta$. He has shown the asymptotic equivalence of Taylor expansions of \eqref{eqn:pdwaic} and harmonic mean \eqref{eqn:hmmicnoM} (without $\bb_i$). From our research, we do see that \eqref{eqn:pdwaic} provides results very close to CV posterior predictive density of each $\by_i\obs$.  This way to look at WAIC also provides the approach to assess statistical significance of differences of WAICs of different models by looking at differences in means of log CV posterior predictive densities, which was advocated by \cite{vehtari2002cv} for CVIC itself. 

% % \subsection{Naive WAIC for Models with Correlated Latent Variables}

For the models given in Section \ref{sec:bmlv} with possibly correlated latent variables, a naive way to approximate CVIC is to apply WAIC directly to the non-integrated predictive density of $\by_{i}\obs$ conditional on $\btheta$ and $\bb_i$:
\begin{equation}
 \label{eqn:pdnwaic}
 \hat{P}\nwaic (\by_i\obs|\by\noi\obs) = \exp\big\{\log (E\postfnoM(P(\by_i\obs|\btheta, \bb_i))) - V\postfnoM(\log (P(\by_i\obs|\btheta, \bb_i)))\big\}. 
\end{equation}
We will refer to \eqref{eqn:pdnwaic} as non-integrated WAIC (or nWAIC for short) method for approximating CV posterior predictive density.  The corresponding information criterion based on \eqref{eqn:pdnwaic} is:
\begin{equation}
 \label{eqn:nwaic}
 \mbox{nWAIC} = -2 \sum_{i=1}^n \log (\hat{P}\nwaic (\by_i\obs|\by\noi\obs)).
\end{equation}
This way to apply WAIC indeed treats latent variables as model parameters. nWAIC is not justified by the theory for WAIC. However,  practitioners may likely apply WAIC to Bayesian models with latent variable this way for the sake of convenience. 

Our research (to be presented next) will show that nWAIC cannot correct for the bias in unit-specific latent variables entirely. We propose to apply WAIC approximation to the integrated predictive density  \eqref{eqn:inpd} to estimate the CV posterior predictive density:
\begin{equation}
 \label{eqn:pdiwaic}
 \hat{P}\iwaic (\by_i\obs|\by\noi\obs) = \exp\big\{\log (E\postfnoM(P(\by_i\obs|\btheta, \bb\noi))) - V\postfnoM(\log (P(\by_i\obs|\btheta, \bb\noi)))\big\}. 
\end{equation}
Accordingly, iWAIC for approximating CVIC is given by :
\begin{equation}
 \label{eqn:iwaic}
 \mbox{iWAIC} = -2 \sum_{i=1}^n \log (\hat{P}\iwaic (\by_i\obs|\by\noi\obs)).
\end{equation}

In Section \ref{sec:iIS}, we have theoretically shown the equivalence of iIS to CV predictive evaluation for models with correlated latent variables, which holds as long as the support of  full data posterior is not a subset of the CV posterior.  However, we haven't proven any sort of equivalences of $\hat{P}\iwaic$ and $\hat{P}\nwaic$ to CVIC.  The derivations of formulae for nWAIC and iWAIC for models with correlated latent variables are only heuristic,  borrowing the asymptotic equivalence of WAIC estimate \eqref{eqn:pdwaic} and CVIC expressed with harmonic mean (IS) (\ref{eqn:hmm-nis}) (without $\bb_i$) for models without latent variables, which is proved by \cite{watanabe2010asymptotic}.

\section{Data Examples} \label{sec:examples}

\subsection{Finite Mixture Models for Galaxy Data} \label{sec:mix}

\begin{figure}[tp]
 \centering

 \caption{Histograms of Galaxy data and three estimated density curves using MCMC samples from fitting finite mixture models with different numbers of components, $K = 4, 5, 6$ and the full data set.}
 \label{fig:histmix}

 \subfloat[$K = 4$]{\includegraphics[width=0.32\textwidth]{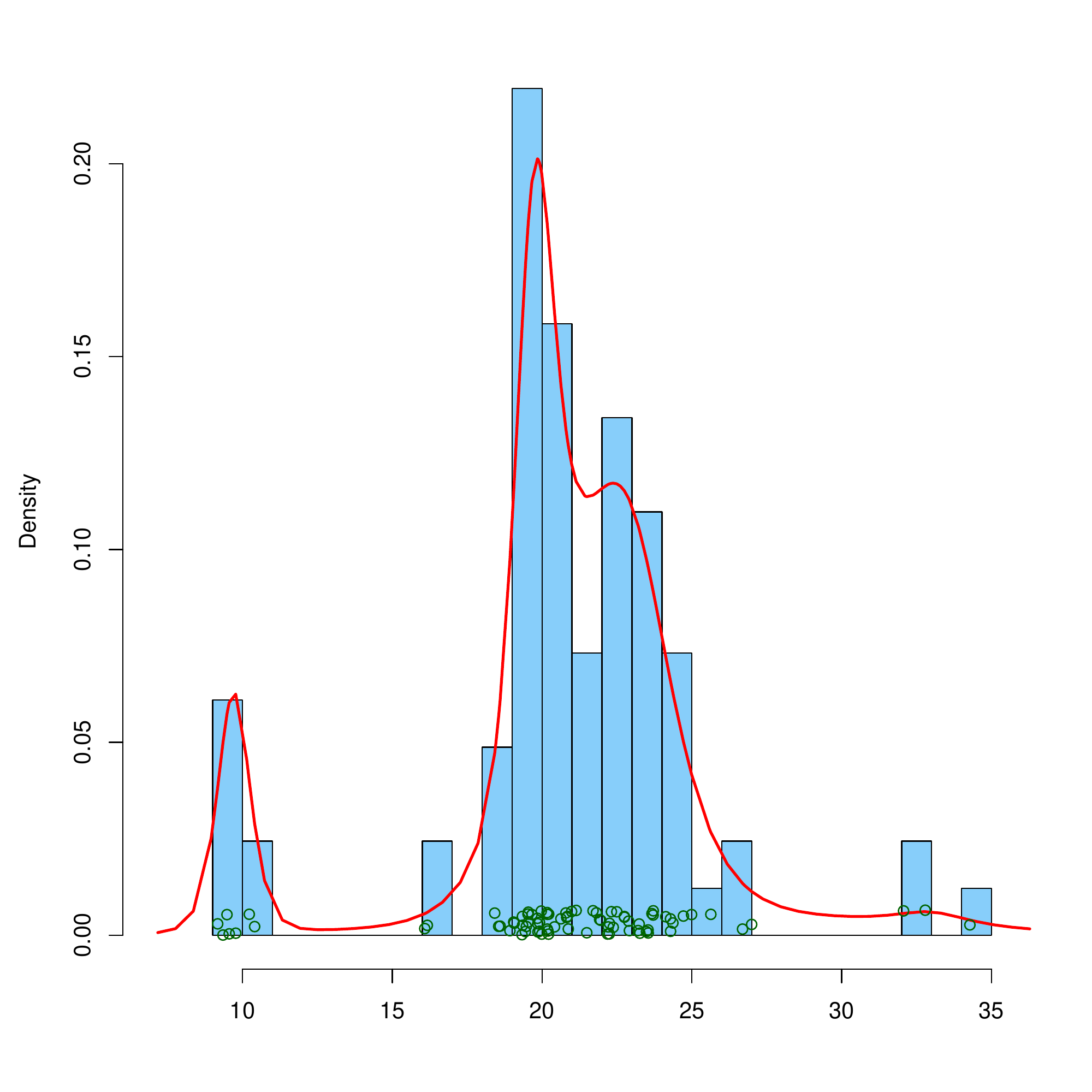} }
 \subfloat[$K = 5$]{\includegraphics[width=0.32\textwidth]{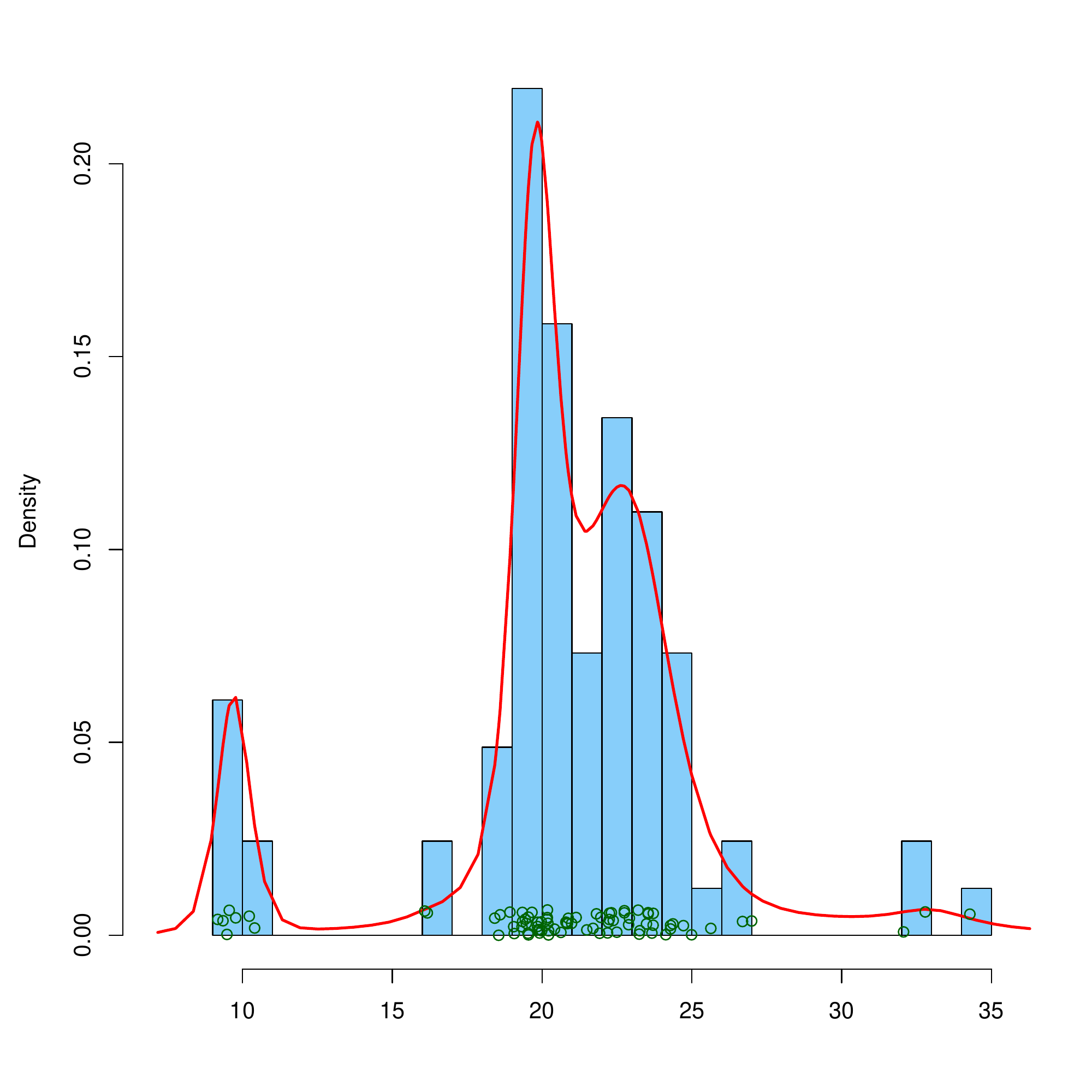} }
 \subfloat[$K = 6$]{\includegraphics[width=0.32\textwidth]{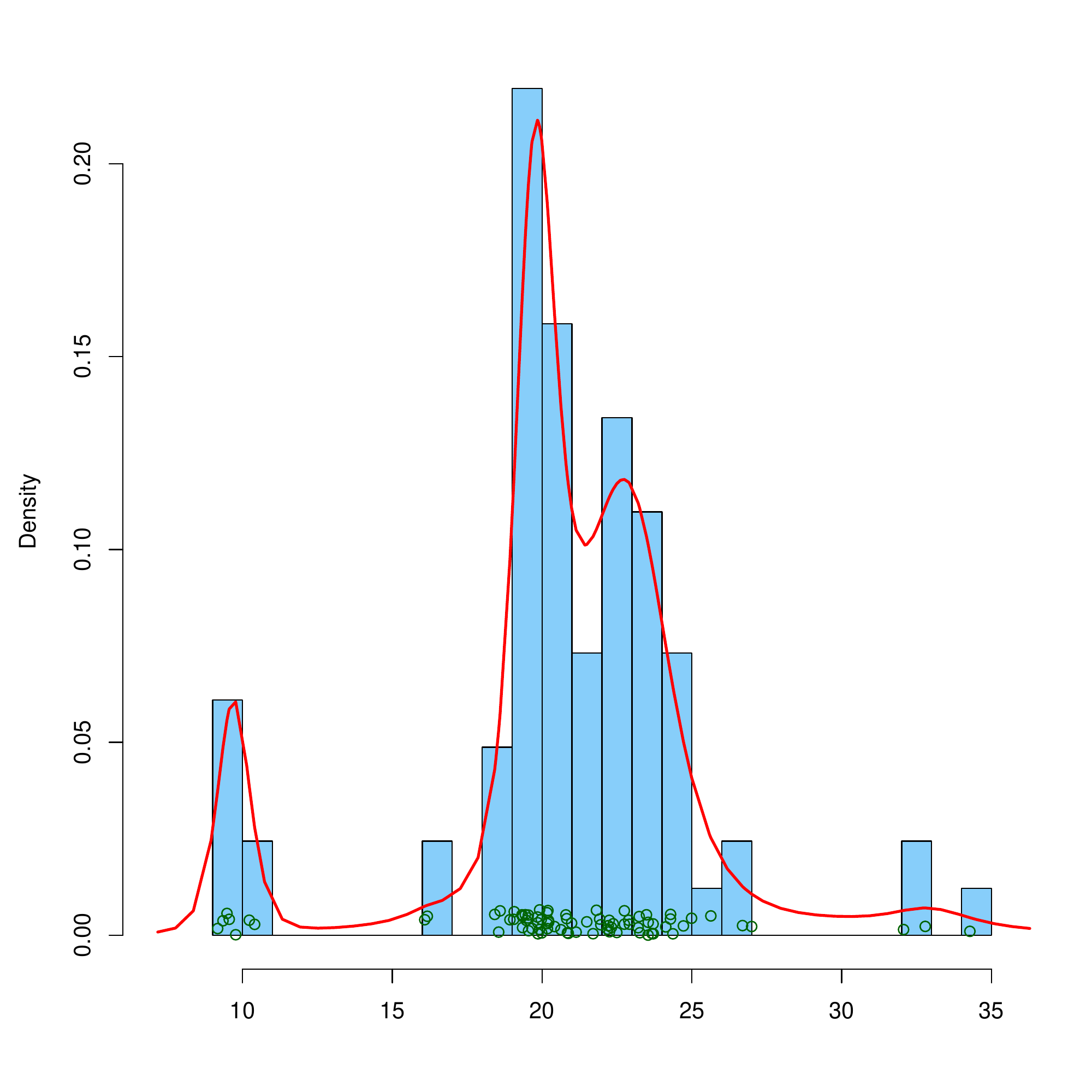} }
 
 \vspace*{-20pt}
\end{figure}

In this section, we look at the performance of iIS and iWAIC in approximating CVIC of fitting finite mixture models to \texttt{Galaxy} data \citep{ postman1986probes, roeder1990density} which is used very often to demonstrate mixture modelling methods. We obtained the data set from R package \texttt{MASS}. The data set is a numeric vector of velocities (km/sec) of 82 galaxies from 6 well-separated conic sections of an ˜unfilled survey of the Corona Borealis region. We applied mixture modelling to the velocities divided by 1000.  A histogram of these 82 numbers is shown in each plot of Figure \ref{fig:histmix}, which  also shows three fitted density functions to be discussed later. Our purpose of computing CVIC for finite mixture models is to determine the numbers of mixture components, $K$, that can adequately capture the heterogeneity in a data but don't overfit the data.  The finite mixture model that we used to fit Galaxy data is as follows:
\begin{eqnarray}
 \label{eqn:mixmodel1}y_i | z_i = k, \bmu_{1:K}, \bsgm_{1:K} &\sim& N(\mu_{k}, \sigma_k^2), \mbox{ for } i = 1,\ldots,n, \\
 z_i |\mb p_{1:K} &\sim& \mbox{Category} (p_1,\ldots, p_K), \mbox{ for } i = 1,\ldots,n, \\
 \mu_k &\sim& N(20, 10^4), \mbox{ for } k = 1, \ldots, K,\\
 \sigma_k^2 &\sim& \mbox{Inverse-Gamma} (0.01, 0.01 \times 20), \mbox{ for } k = 1, \ldots, K, \\
 \label{eqn:mixmodel2}p_k &\sim& \mbox{Dirichlet} (1, \ldots, 1) \mbox{ for } k = 1, \ldots, K.
 \end{eqnarray}
Here we set the prior mean of $\mu_k$ to 20, which is the mean of the 82 numbers, and set the scale for Inverse Gamma prior for $\sigma_k^2$ to 20, which is the variance of the 82 numbers. 

The finite mixture model (equations \eqref{eqn:mixmodel1} - \eqref{eqn:mixmodel2}) falls in the class of models depicted by Figure \ref{fig:blvm}: the observed variable is $y_i$, the model parameters $\btheta$ is $(\bmu_{1:K}, \bsgm^2_{1:K}, \mb p_{1:K})$, and the latent variable $\bb_i$ is mixture component indicator $z_i$. In this model,  the latent variables $z_1, \ldots, z_n$ are independent given the model parameter $\btheta$. It follows that $y_1,\ldots, y_n$ are independent given $\btheta$. 
 
We used JAGS \citep{plummer2003jags} to run MCMC simulations for fitting the above model  to \texttt{Galaxy} data with various choice of $K$. To avoid the problem that MCMC may get stuck in a model with only one component, we followed JAGS \texttt{eyes} example to restrict the MCMC to have at least a data point in each component. All MCMC simulations started with a randomly generated $\mb z_{1:n}$, and ran 5 parallel chains, each doing 2000, 2000, and 100,000  iterations for adapting,  burning, and sampling, respectively. 

We ran actual 82 cross-validatory MCMC simulations with each of the 82 numbers removed (set to \texttt{NA} in JAGS). After each simulation, we computed actual CV posterior predictive density $P(y_i\obs|\by\noi\obs)$ using equation \eqref{eqn:cvevalnoM} with evaluation function $a$ set to $\phi(y_{i}\obs | \mu_{z_{i}}, \sigma^{2}_{z_{i}})$, where $\phi$ represents normal density. Using all 82 values of CV posterior predictive densities, we can compute CVIC using equation \eqref{eqn:cvic}. The CVICs for different choices of $K$ based on one simulation for each $K$ are displayed in Table \ref{tbl:mixic}.  We repeated computing CVICs quite a few times, and the results were almost the same, with only differences in the 2nd decimal.  

We then considered approximating CVIC using four different methods (nIS, nWAIC, iIS, iWAIC) from a single MCMC simulation that is based on all of the 82 numbers. The non-integrated predictive density for this model is $P(y_i\obs|z_i, \btheta)$ as specified in \eqref{eqn:mixmodel1}; this is normal density with mean $\mu_{z_i}$ and standard deviation $\sigma_{z_i}$, denoted by $\phi(y_i\obs|\mu_{z_i}, \sigma_{z_i})$. The values of $P(y_i\obs|z_i, \btheta)$ computed with a collection of MCMC samples of $(z_i,\btheta)$ are then used for computing nIS and nWAIC approximates  of CV posterior predictive densities (with equations \eqref{eqn:hmmicnoM} and \eqref{eqn:pdnwaic} respectively). We can then compute nIS information criterion and nWAIC by plugging the approximates of CV posterior predictive densities into \eqref{eqn:cvic}. The integrated predictive density is $P(y_i\obs|\btheta) = \sum_{k=1}^K p_k\phi(y_i\obs|\mu_k,\sigma_k)$ ( note that $\mb z_{-i}$ and $y_i$ are independent given $\btheta$). We can then use $P(y_i\obs|\btheta)$ for computing iIS and iWAIC approximates of CV posterior predictive densities (with equations \eqref{eqn:pdhmmM} and \eqref{eqn:pdiwaic} respectively), and corresponding information criterion values. In this example, iIS and iWAIC are just applications of IS and WAIC to mixture models with  latent variables $\mb z\wi$ integrated out. 

% latex table generated in R 2.15.3 by xtable 1.7-1 package
% Wed Sep  4 14:44:48 2013
\begin{table}[htp]

\centering

\caption{Comparisons of 5 information criteria for mixture models. The numbers are the averages of ICs from 100 independent MCMC simulations. The numbers in brackets indicates standard deviations.} \label{tbl:mixic}

\begin{tabular}{rllllll}
  \hline
$K$ & DIC & nWAIC & nIS & iWAIC & iIS & CVIC \\ 
  \hline
  2 & 445.38(1.64) & 420.27(0.39) & 425.63(3.45) & 449.56(0.14) & 449.62(0.17) & 450.55 \\ 
    3 & 528.78(45.12) & 384.94(9.94) & 391.29(6.17) & 437.23(4.70) & 436.43(3.79) & 427.46 \\ 
    4 & 774.85(31.58) & 339.91(1.87) & 363.55(5.32) & 422.43(0.53) & 422.76(0.54) & 423.16 \\ 
    5 & 710.88(25.34) & 328.19(0.29) & 362.30(3.70) & 421.02(0.09) & 421.41(0.10) & 421.10 \\ 
    6 & 679.95(17.48) & 323.62(1.33) & 355.49(5.72) & 420.97(0.27) & 421.35(0.31) & 421.34 \\ 
    7 & 675.27(18.57) & 321.61(0.30) & 364.41(4.49) & 421.25(0.07) & 421.64(0.12) & 421.53 \\ 
   \hline
\end{tabular}
\end{table}

For each choice of $K$, we computed the above four criteria as well as DIC (using R package R2jags) for 100 independent MCMC simulations. Table \ref{tbl:mixic} shows the means of these 100 information criterion values for each approximation method, with standard deviations shown in brackets. From the table, we see that the naive applications of WAIC and IS to non-integrated predictive densities $P(y_i\obs|z_i, \btheta)$ do not work satisfactorily. They are both highly downward biased. Furthermore, nWAIC chooses over-complex models because nWAICs keep decreasing until $K=7$, and nIS estimates of CVIC have very high variances. DICs for this example turn into a mess because the model parameters are non-identifiable. 
iIS and iWAIC provide significantly closer estimates of actual CVIC, with much smaller standard deviations, than other methods. These results show that using integrated predictive densities significantly improves accuracy of nIS and nWAIC. The results of iWAIC may not be surprising because here iWAIC is just application of WAIC to the marginalized models with latent variables $\mb z_{1:n}$ integrated out, in which observed variables $y_1,\ldots, y_n$ are independent given model parameters. \citet{watanabe2010asymptotic} has proven the asymptotic equivalence of WAIC and CVIC in such models.  iIS is also theoretically justified in Section \ref{sec:iIS}.

% latex table generated in R 2.15.3 by xtable 1.7-1 package
% Wed Sep  4 16:57:24 2013
\begin{table}[ht]

\caption{One-sided paired t-test p-values for comparing means of 82 log posterior predictive densities for Galaxy data given by mixture models with different number of mixture components, $K$.} \label{tbl:mixpv}

\centering
\begin{tabular}{crrrrr}
  \hline
 pair of models & nWAIC & nIS & iWAIC & iIS & CVIC \\ 
  \hline
  $K=$3 vs $K=$ 2 & 0.000 & 0.000 & 0.016 & 0.013 & 0.010 \\ 
  $K=$ 4 vs $K=$ 3 & 0.000 & 0.019 & 0.030 & 0.032 & 0.190 \\ 
  $K=$ 5 vs $K=$ 4 & 0.000 & 0.249 & 0.070 & 0.066 & 0.027 \\ 
  $K=$ 6 vs $K=$ 5 & 0.002 & 0.203 & 0.489 & 0.476 & 0.674 \\ 
  $K=$ 7 vs $K=$ 6 & 0.110 & 0.840 & 0.716 & 0.711 & 0.700 \\ 
   \hline
\end{tabular}
\end{table}

CVIC is the sum of minus twice of log CV posterior predictive densities. Therefore, the statistical significance of the differences of two CVICs (or estimates) can be accessed by looking at the population mean differences of two groups of log CV posterior predictive densities \citep{vehtari2001bayesian, vehtari2002cv}. We conducted one-sided paired t-test to test whether a finite mixture model with $K$ components provides a better fit (larger mean of CV posterior predictive densities) to \texttt{Galaxy} data than a mixture model with $K-1$ components. The p-values of the comparisons for $K = 3, \ldots, 7$ for actual CV posterior predictive densities are given in Table \ref{tbl:mixpv} (column CVIC). We also conducted the same test for log CV posterior predictive densities estimated by four different methods (nIS, iIS, nWAIC, iWAIC). Due to the variations in these estimates, we computed the p-values 1000 times by randomly drawing two simulation results from models with $K$ and $K-1$ components. We then computed the mean of the 1000 p-values. Table \ref{tbl:mixpv} shows the results for all four different estimation methods. From the table, we see that iIS and iWAIC provides much closer p-values to those based on actual CV posterior predictive densities than nIS and nWAIC. These p-values indicates that mixture models with 5 components are adequate to capture the heterogeneity in Galaxy data, and 6-component mixture models does not provide better fit with statistical significance. These conclusions can be visualized by the density curves given by fitting resulting with $K=4,5,6$, where the curves with $K=4$ and $K=5$ are  different, but the curves with $K=5$ and $K=6$ are almost the same.

\begin{figure}[htp]
 \centering
 
\caption{Scatter-plot of non-integrated predictive densities against $\mu_{z_i}$, given MCMC samples from the full data posterior (\ref{fig:nint-full}) and the actual CV posterior with the 3rd number removed (\ref{fig:nint-cv}), when $K = 5$ components are used.} \label{fig:nint}
 
 \subfloat[]
          {\label{fig:nint-full}\includegraphics[width=0.5\textwidth]{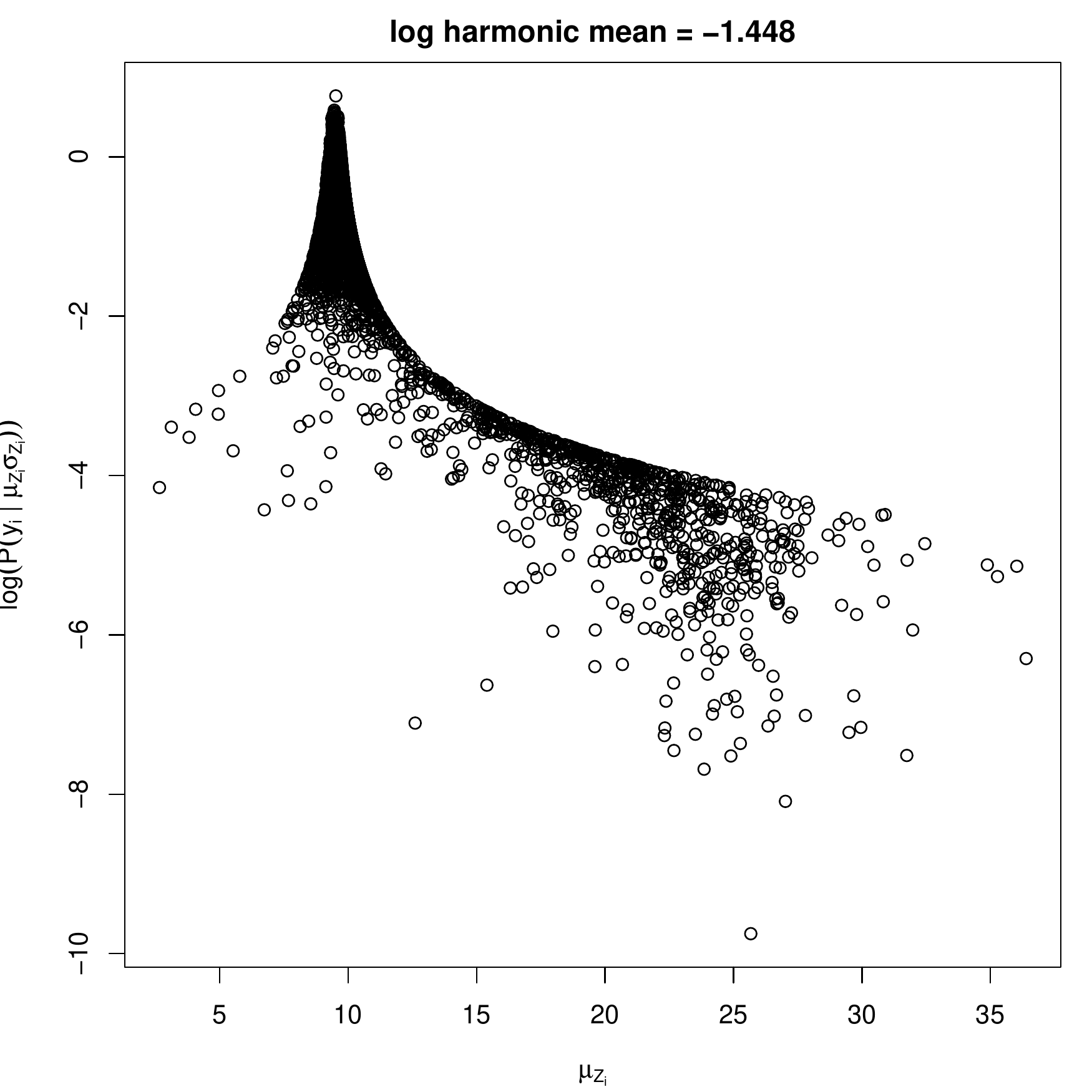}}
 \subfloat[]
          {\label{fig:nint-cv}\includegraphics[width=0.5\textwidth]{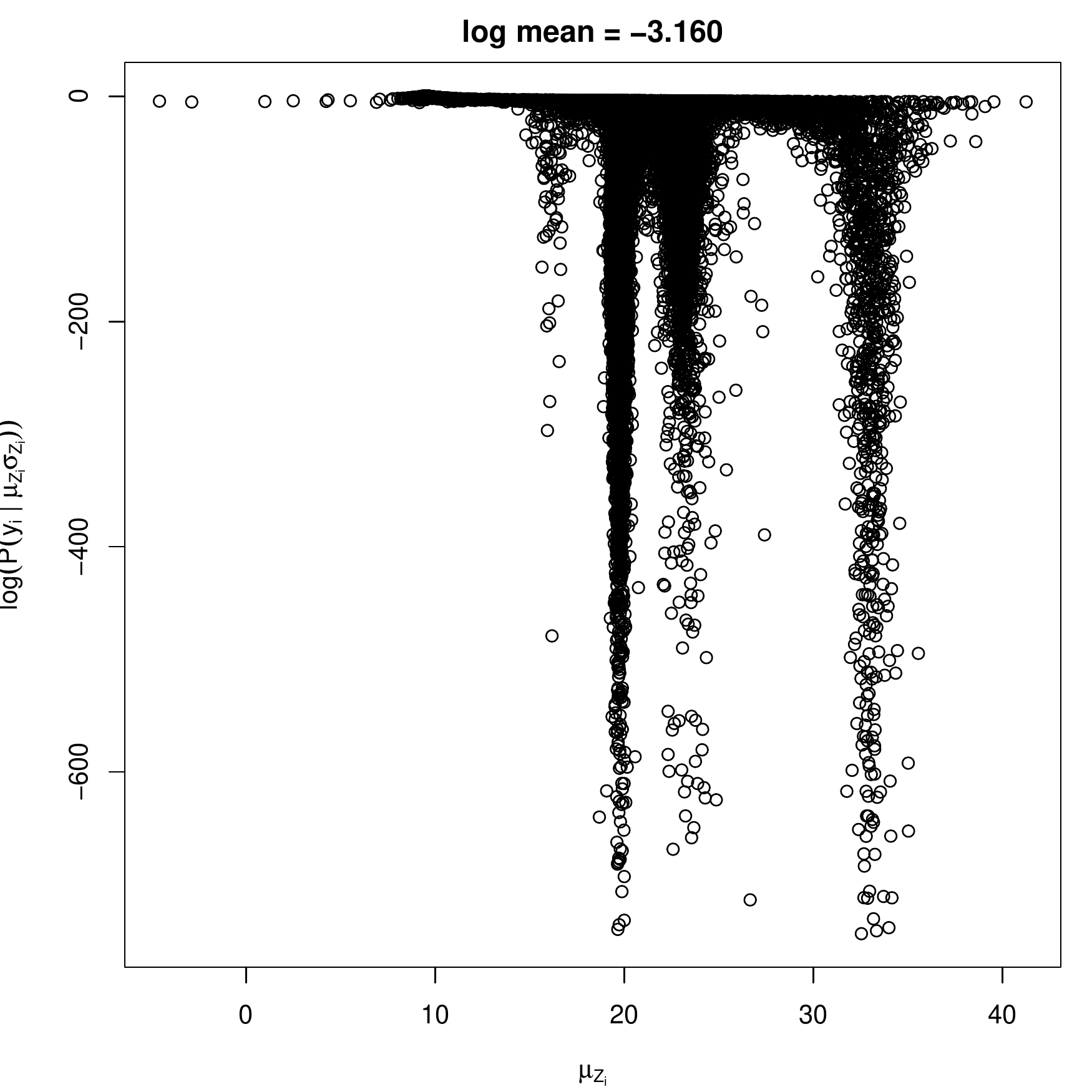}}

\vspace*{-10pt}

\end{figure}

Last, we explain why naive applications of IS and WAIC to non-integrated predictive densities cannot provide good estimates of CV posterior predictive densities. Figure \ref{fig:nint} show scatter-plots of the log non-integrated predictive density, $\log(P(y_i\obs|z_i, \btheta))= \log(\phi(y_i\obs| \mu_{z_i}, \sigma_{z_i}))$, against $\mu_{z_i}$, computed with each MCMC sample of $(z_i, \bmu_{1:K}, \bsgm_{1:K})$ from the full data posterior (Figure \eqref{fig:nint-full}) and the actual CV posterior with the $y_i\obs$ removed (Figure \eqref{fig:nint-cv}), where $y_i\obs$ is the 3rd of the 82 numbers. From the figure, we see great discrepancy between the posterior distribution of the non-integrated predictive density with and without $y_i\obs$ included in MCMC simulations. When we simulate MCMC with the full data ($y\obs_i$ included), most of the $z_i$ visit components that fit $y_i\obs$ well, with most of $\mu_{z_i}$ are around 10. Thus, the non-integrated predictive densities are mostly very high. When we simulate MCMC with $y_i\obs$ removed, most of the $z_i$ visit large components, hence the $\mu_{z_i}$ visits much more often the interval from 10 to 35 that do not fit $y\obs_i$ well. The reason is that without the inclusion of $y_{i}\obs$, the $z_{i}$ will more likely take larger components. Thus, values of $P(y_{i}|\btheta,z_{i})$ in the CV posterior are very low, with greatly lower order in magnitude than in the full data posterior. This indicates that the difference between the CV posterior  and full data posterior of $z_{i}$ is huge.  Applying IS and WAIC to the non-integrated predictive densities alone is unable to correct for much of the bias due to the inclusion of $y_i\obs$ in MCMC simulation. By averaging the non-integrated predictive density over regenerated $z_i$ given $\btheta$ but not $y_i\obs$, we significantly reduce  the optimistic bias in $P(y_i\obs|\btheta, z_i$) due to inclusion of $y_i\obs$. This explains why iIS and iWAIC provide significantly closer estimates to CVIC than nIS and nWAIC.  

\subsection{A Simulation Study with Finite Mixture Models} \label{sec:simmix}

\begin{figure}[htp]

\caption{Kernel Density of a Simulated Data Set.} \label{fig:simy}
\vspace*{-10pt}
\begin{center}
\includegraphics*[width = 0.7\textwidth, height = 0.3\textheight, trim = 0 0 0 25]{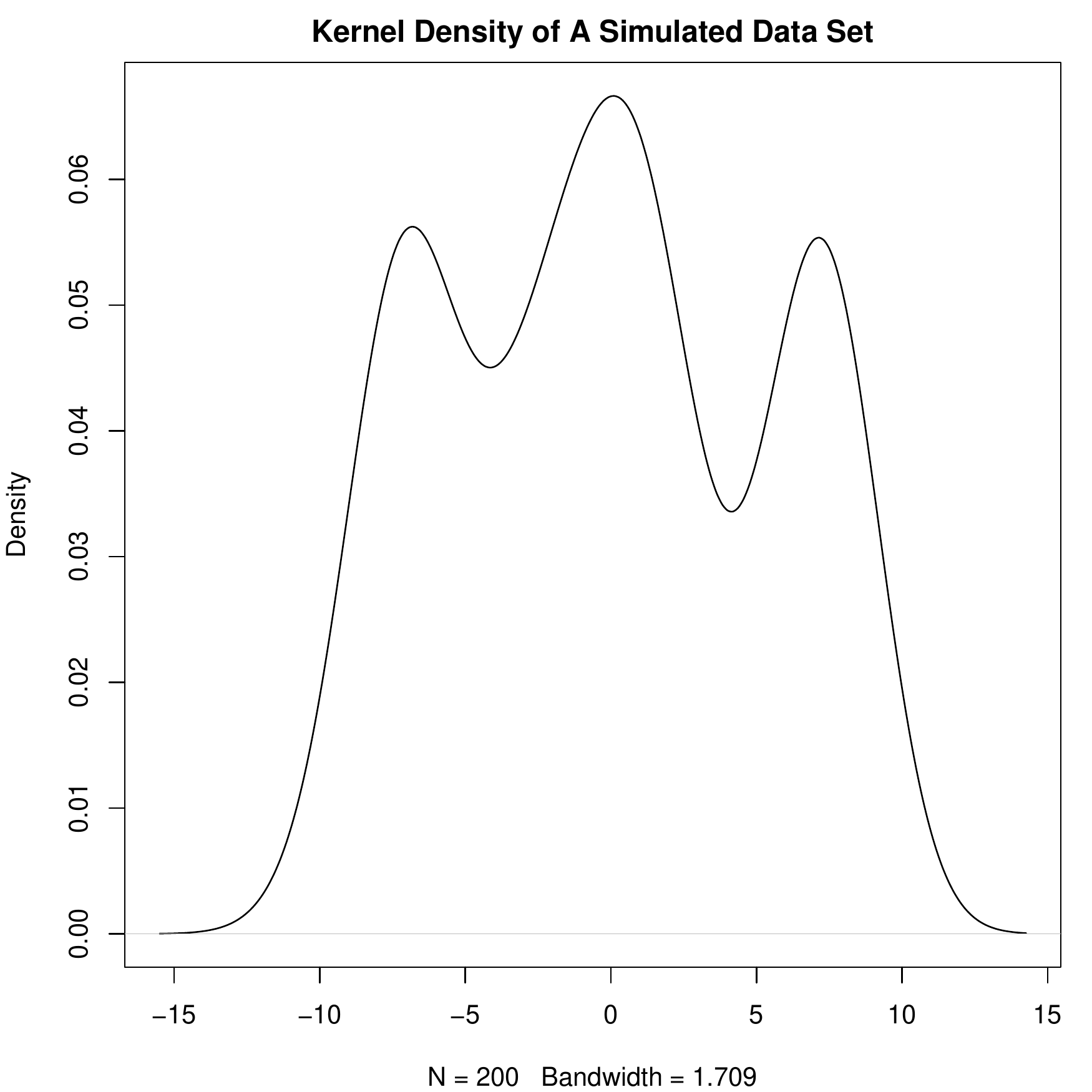}
\end{center}
\vspace*{-20pt}
\end{figure}

In this section, we report a simulation study with the same mixture models described in Section \ref{sec:mix}. We simulated 100 data sets, each containing 200 data points $y_{i}$ from  a mixture distribution with $K = 4$ normal components: $(1/4) N(-7, 1) + (1/4) N(-2,1) + (1/4) N(1,1) + (1/4) N(7,1)$.  The kernel density of one of the data sets are shown in Figure \ref{fig:simy}. From this plot, we see that the middle two components may be hard to separate in some data sets. 

We fitted each of the 100 data sets using the exactly same way described in Section \ref{sec:mix}, and then computed information criterion (IC) using each of the five methods (nWAIC, nIS, iWAIC, iIS and DIC).  Table \ref{tab:twoICsimmix} show the IC values for two selected data sets. Table \eqref{tab:simmix_meanIC} shows averages of IC values in 100 data sets, for each model indexed by $K$ (row) and for each method for approximating CVIC (column). Table \eqref{tab:simmix_freqsel} shows frequencies of selected models in 100 data sets  by looking at the minimum IC value computed with each of the five methods (column).  

\begin{table}[htp]
\caption{Values of information criterion for two selected data sets. The bold-faced numbers show the smallest IC values for each method.} \label{tab:twoICsimmix}
\small 
\subfloat[iIS and iWAIC select $K=4$\label{tab:simmix_ic1}]{
\centering
\begin{tabular}{rrrrrr} %data set #2
  \hline
 $K$& nWAIC & nIS & iWAIC & iIS & DIC \\ 
  \hline
2 & 1022.95 & 1081.11 & 1163.96 & 1163.99 & 1143.19 \\ 
  3 & 690.39 & 855.85 & 1088.20 & 1088.26 & \textbf{850.28} \\ 
  4 & 642.82 & 782.94 & \textbf{1083.16} & \textbf{1083.28} & 1416.81 \\ 
  5 & 640.27 & 754.13 & 1084.29 & 1084.48 & 1351.18 \\ 
  6 & 638.25 & 756.82 & 1085.25 & 1085.51 & 1382.83 \\ 
  7 & \textbf{637.12} & \textbf{727.84} & 1086.46 & 1086.76 & 1479.03 \\ 
   \hline
\end{tabular}
}
\subfloat[iIS and iWAIC do not select $K=4$\label{tab:simmix_ic2}]{
\begin{tabular}{rrrrrr} %data set #62
  \hline
 $K$& nWAIC & nIS & iWAIC & iIS & DIC \\ 
  \hline
2 & 1144.14 & 1122.07 & 1202.25 & 1199.90 & 1192.42 \\ 
  3 & 758.35 & 924.19 & 1101.48 & 1101.53 & \textbf{1024.83} \\ 
  4 & 706.30 & 830.63 & 1095.42 & 1095.54 & 1510.18 \\ 
  5 & 691.02 & 821.51 & 1094.84 & 1094.99 & 1561.51 \\ 
  6 & 679.71 & 800.38 & \textbf{1094.64} & \textbf{1094.80} & 1652.05 \\ 
  7 & \textbf{673.63} & \textbf{794.05} & 1094.69 & 1094.87 & 1740.80 \\ 
   \hline
\end{tabular}
}
\end{table}
\begin{table}[ht]
\caption{Model selection results with 100 data sets simulated from finite mixture models with $K=4$. Table \eqref{tab:simmix_meanIC} shows averages of IC values in 100 data sets, for each model with different $K$ components (row) and computed with different method (column). Each column of Table \eqref{tab:simmix_freqsel} shows frequencies of selected models indexed by $K$ by looking at the minimum IC value in 100 data sets.}
\centering
\subfloat[Average of IC values in 100 data sets\label{tab:simmix_meanIC}]{
\begin{tabular}{rrrrrr}
  \hline
  $K$ & nIS & nWAIC & iIS & iWAIC & DIC \\ 
  \hline
 2 & 1112.48 & 1103.95 & 1181.60 & 1182.33 & 1248.97 \\ 
  3 & 922.88 & 751.58 & 1105.18 & 1105.11 & 990.51 \\ 
  4 & 827.06 & 682.62 & 1099.42 & 1099.26 & 1572.80 \\ 
  5 & 810.42 & 674.39 & 1099.18 & 1098.96 & 1562.05 \\ 
  6 & 801.24 & 669.57 & 1099.60 & 1099.31 & 1630.02 \\ 
  7 & 796.65 & 666.39 & 1100.09 & 1099.77 & 1700.12 \\ 
   \hline
\end{tabular}
} ~~~
\subfloat[Frequencies of selected models\label{tab:simmix_freqsel}]{
\begin{tabular}{rrrrrr}
  \hline
$K$ & nIS & nWAIC & iIS & iWAIC & DIC \\ 
  \hline
2 & 0 & 0 & 0 & 0 & 2 \\ 
  3 & 0 & 0 & 15 & 15 & 94 \\ 
  4 & 6 & 15 & 39 & 37 & 4 \\ 
  5 & 10 & 4 & 21 & 20 & 0 \\ 
  6 & 30 & 8 & 11 & 13 & 0 \\ 
  7 & 54 & 73 & 14 & 15 & 0 \\
  \hline
\end{tabular}

}
\end{table}

In both of the data sets shown in Table \ref{tab:twoICsimmix}, nWAIC and nIS select the  model with $K=7$, which is more flexible than the true data generating model, which has $K=4$.  This is typical for the 100 data sets, which can be seen from Tables \eqref{tab:simmix_meanIC} and \eqref{tab:simmix_freqsel}. DIC, on the other hand, almost always selects the model with $K=3$ which is simpler than the true model ($K=4$).  These results are consistent with what we observe from the analysis of \texttt{Galaxy} data. For iWAIC and iIS, their IC values have a sharp decrease from $K=2$ until $K=4$, compared to the changes in the IC values after $K=4$, which stabilize and have only very small variation. This small variation sometimes leads to wrong model selection results if one chooses the model with the smallest IC value, for example in the data set shown by Table \eqref{tab:simmix_ic2}.  Therefore,  IC may not be sensitive enough in penalizing over-complex models.  Overall, in this example, iWAIC and iIS outperform nWAIC, nIS and DIC in comparing models because of the improved estimation of CVIC. With IC values computed by iWAIC and iIS, the appropriate decision ($K=4$) can be made for most data sets if one does not simply look at which model has the smallest IC value but also the change of IC values in all models considered.

In this example, IC cannot penalize over-complex models sensitively.  The insensitivity occurs because the posterior inference with MCMC itself is robust to over-complexity in models, that is, MCMC simulation can automatically adjust the model complexity. For example, in this example, although we fit a mixture model with $K=7$ components, some components have very small proportions in MCMC samples, which is effectively a simpler model. This has been long known as a good property for Bayesian inference, see extensive discussions by \citet{neal1995bayesian}. However, this poses difficulty in model selection by looking at CVIC. We've noticed that recently  \citet{wang2014difficulty} have also discussed the insensitivity of CVIC. They explain the insensitive as that the the criterion CVIC itself is not sensitive in distinguishing models for binary data.  Overall,  how to determine a threshold for CVIC (even computed with actual cross-validation) for selecting models, particularly among models with slight difference, is still a problem, which demands further study. Looking at the population mean of log CV posterior predictive density may be an option, as we discuss in \texttt{Galaxy} data analysis results. However, we feel that generally this may be too conservative because a model is better than another not because it can provides better predictive accuracy for ``most''  observations (resulting in a sharp change in population mean --- CVIC), but rather it can provide better prediction only for a fraction of observations.  Perhaps, we should look at the proportion of units whose predictions have been improved with a more complex model.

\subsection{Random Spatial Effect Models for Scottish Lip Cancer Data} \label{sec:spatial}

%Instructions how to write this part:

%In this section, we compare different information criteria with actual CVIC for 4 different models for london suicide data. describe the data set more ...

In this section, we investigated the performance of iIS and iWAIC in an analysis of Scottish lip cancer data, which was used in \citet{stern2000posterior, spiegelhalter2002bayesian, plummer2008penalized}.  The data set was extracted from the paper by \citet{stern2000posterior}.  The data represents male lip cancer counts (over the period 1975 - 1980) in the $n = 56$ districts of Scotland.  At each district $i$,  the data include these fields: 
\begin{inparaenum}[(1)]
\item the number of observed cases of lip cancer, $y_{i}$; 
\item the number of expected cases, $E_{i}$,  calculated based on standardization of  ``population at risk" across different age groups;
\item the percent of population employed in agriculture, fishing and forestry, $x_{i}$, used as a covariate; and 
\item a list of the neighbouring regions. 
\end{inparaenum}

The $y_i$, for $i = 1,\ldots,n$, is modelled as an independent Poisson random variable conditional on $\lambda_{i}$ and $E_{i}$:
\begin{equation}
 y_i |E_{i}, \lambda_{i} \sim \mbox{Poisson}(\lambda_i E_i),
\end{equation} 
where $\lambda_i$ denotes the underlying relative risk for district $i$, and $E_{i}$ stands for expected counts. Let $s_{i} = \log (\lambda_{i})$ and $\mb X = (x_{1}, \ldots, x_{n})'$. We consider four different models for the vector $\mb s = (s_{1}, \cdots, s_{n})'$ conditional on $\mb X$ and neighbouring information between districts: \begin{eqnarray}
&&\mbox{spatial+linear (called \textit{full} for short)}: \mb s  \sim N_{n}(\alpha + \mb X\beta,  \Phi\tau^{2}), \label{eqn:ldmodel1} \\
&&\mbox{spatial}:  \mb s  \sim N_{n}(\alpha,  \Phi\tau^{2}),      \label{eqn:ldmodel2} \\
&&\mbox{linear}: \mb s  \sim N_{n}(\alpha + \mb X\beta,  I_{n}\tau^{2}),     \label{eqn:ldmodel3}   
\\
&&\mbox{exchangable}: \label{eqn:ldmodel4} \mb s  \sim N_{n}(\alpha,  I_{n}\tau^{2}),            
\end{eqnarray} 
where  $\Phi = (I_{n} - \phi C)^{-1}M$ is  a matrix for capturing the spatial correlations amongst the $n$ districts, in which, the elements of $C$ are: $c_{ij}= (E_{j}/E_{i})^{1/2}$ if areas $i$ and $j$ are neighbours, and 0 otherwise; the elements of $M$ are: $m_{ii}=E_{i}^{-1}$ and $m_{ij}=0$ if $i\not=j$. The multivariate normal distributions with $\Phi$ as covariance matrix are called \textbf{proper conditional auto-regression (CAR) model}.  Derived from the joint distribution in \eqref{eqn:ldmodel1}, the conditional distribution of $s_{i}|\mb s_{-i}, \alpha, \beta, \phi$ is: 
\begin{equation}
s_{i}|\mb s_{-i}, \btheta \sim N(\alpha + x_{i}\beta + \phi\sum_{j\in N_{i}}(c_{ij}(s_{j}-\alpha - x_{j}\beta)), \tau^{2} m_{ii}), 
\label{eqn:pcarconds}
\end{equation}
where $N_{i}$ is the set of neighbours of district $i$. From \eqref{eqn:pcarconds}, we see that $\phi$ controls the degree of spatial dependency of $s_{i}$ on its neighbours.  At a higher level, diffused priors are assigned to $\alpha, \beta, \tau$, and $\phi$:
$
\alpha \sim N(0, 1000^{2}),
$
$
\beta \sim N(0,  1000^{2}),
$
$
\tau^{2} \sim \mbox{Inv-Gamma} (0.5, 0.0005),
$
$
\phi \sim \mbox{Unif}(\phi_0, \phi_{1}),
$
where $(\phi_{0}, \phi_{1})$ is the interval for $\phi$ such that $\Phi$ is positive-definite \citep[see][]{stern2000posterior}.  In model \eqref{eqn:ldmodel1}, we consider both spatial and linear effects of $x_{i}$ in modelling $\mb s$. One may also consider other models. Model \eqref{eqn:ldmodel2} considers only spatial effect; model \eqref{eqn:ldmodel3} considers only linear effect; and model \eqref{eqn:ldmodel4} considers none of spatial and linear effect.  We are interested in comparing goodness-of-fits of the four models to lip cancer data set so as to determine which model is the most appropriate for Scottish lip cancer data.  CVIC is one criterion for measuring goodness-of-fit. 

All the above four models belong to the class of Bayesian latent variable models depicted by Figure \ref{fig:blvm}. The observable variable is $y_i$, the latent variable is $s_i$ , and the model parameters $\btheta$ in model \eqref{eqn:ldmodel1} are $(\alpha, \beta, \tau, \phi)$, and a subset of it for other models depending on which are used in respective models.  We used OpenBUGS through R package \texttt{R2OpenBUGS} to run MCMC simulations for fitting each of the above models to Scottish lip cancer data. For each simulation, we ran two parrallel chains, each with 15000 iterations, and the first 5000 were discarded as burn-in.  

For each model, we first ran actual 56 cross-validatory MCMC simulations with each of the 56 obervations removed (set to \texttt{NA} in OpenBUGS) and then computed actual CV posterior predictive density $P(y_i\obs|y_{-i}\obs)$ using equation \eqref{eqn:cvevalnoM} with evaluation function set to $\mbox{dpoisson}(y_{i}\obs|\lambda_{i} E_{i})$ --- Poisson probability mass function with parameter $\lambda_{i}E_{i}$. Then we computed CVIC using equation \eqref{eqn:cvic}. We computed actual CVIC 10 times for each model although actual LOOCV gives very stable results. The averages and standard deviations of 10 CVICs for different models are displayed in Table \ref{tbl:lipcancer}. From this table, we see that the spatial+linear model is optimal for the Scottish lip cancer data according to CVIC.  

We then consider approximating CVIC with four different methods (nIS, nWAIC, iIS, and iWAIC) from a single MCMC simulation based on all of the 56 observations.  The non-integrated predictive density used in computing nIS and nWAIC with equations \eqref{eqn:hmmicnoM} and \eqref{eqn:pdnwaic} is $\mbox{dpoisson}(y_{i}\obs|\lambda_{i}E_{i})$, where $\lambda_{i}=\exp (s_{i})$.  Next, we describe how to compute  iIS and iWAIC for model \eqref{eqn:ldmodel1}.  The integrated predictive density \eqref{eqn:inpd} required by \eqref{eqn:pdhmmM} and \eqref{eqn:pdiwaic}  is:
\begin{equation}
P(y_i\obs\,|\, \btheta, \mb s\noi) = \int  \mbox{dpoisson}(y_{i}\obs|\lambda_{i}E_{i}) P(s_i\, | \, \btheta, \mb s\noi)d s_i,
\label{eqn:intpdkriging}
\end{equation}
where $P(s_{i}|\btheta, \mb s\noi)$ is given by equation \eqref{eqn:pcarconds}. Because there is no closed form for the integral \eqref{eqn:intpdkriging}, we use Monte Carlo method  to estimate it by generating 200 random numbers from $P(s_{i} | \mb s\noi, \btheta)$ (note, this is done for each retained MCMC sample of $(\btheta, \mb s\wi)$ and each validation unit $i$, with $s_{i}$ alternately discarded).   Finally,  based on computed values of $P(y_i\obs|\btheta, \mb s\noi)$ for all MCMC samples, we can then compute iIS and iWAIC approximates of CV posterior predictive densities (with equations \eqref{eqn:pdhmmM} and \eqref{eqn:pdiwaic} respectively) and then corresponding iIS information criterion and iWAIC.  iIS and iWAIC are computed similarly for models \eqref{eqn:ldmodel2} - \eqref{eqn:ldmodel4}, with only a change of the conditional distribution \eqref{eqn:pcarconds} according to their joint prior distributions.

We repeated computing the values of the above four criteria as well as DIC for 100 independent MCMC simulations based on each model. The means of these 100 information criterion values for each method and each model are shown in Table \ref{tbl:lipcancer}, with standard deviations shown in brackets.  We see that,  iIS and iWAIC provide significantly closer approximates to actual CVIC than nIS, nWAIC and DIC; furthermore, the approximates by iWAIC and iIS are almost identical to actual CVIC. In contrast, DIC has large biases and variances when spatial effects are considered, and also the mean DIC of the spatial + linear model is bigger than the mean DIC of the model with spatial effects only. This suggests that,  if we randomly draw one MCMC simulation out of the 100 ones based on each model, the probability that DIC does NOT pick up the spatial+linear model as the optimal model is high (56.6\% if we assume the DICs are normally distributed). nWAIC and nIS also have large biases and variances. In particular, nWAIC nearly never chooses the spatial+linear model (with a probability close to 1 if nWAICs are normally distributed). nIS has a good chance (0.92 if the values are normally distribute) to choose the spatial+linear model. However,  nIS is numerically unstable, with fairly large variance, which has been well-known to many people \citep{spiegelhalter2002bayesian}. In summary, the integration applied to latent variables associated with each validation unit substantially improve the estimates of CVIC given by nWAIC and nIS. 

% !TEX root =  iis.tex

\begin{table}[t]
\caption{Comparisons of information criteria for Scottish lip cancer data. Except for CVIC, each table entry shows the average of 100 information criterion values computed from 100 independent MCMC simulations, and the standard deviation in bracket. For CVIC, the average and standard deviation are from 10 independent LOOCV evaluations. } \label{tbl:lipcancer}
\centering
\begin{tabular}{rllllll}
  \hline
 & DIC & nWAIC & nIS & iWAIC & iIS & CVIC \\ 
  \hline
spa.+lin. & 269.43(12.30) & 306.82(0.21) & 335.54(1.27) & 344.47(0.12) & 345.21(0.19) & 343.88(0.14) \\ 
  spatial & 266.79(10.15) & 304.61(0.18) & 338.77(1.85) & 354.11(0.06) & 356.06(0.37) & 352.54(0.14) \\ 
  linear & 310.42(0.11) & 306.94(0.21) & 338.81(3.02) & 350.48(0.05) & 350.54(0.05) & 349.48(0.11) \\ 
  exch. & 312.57(0.12) & 306.74(0.17) & 346.55(3.46) & 368.01(0.03) & 368.08(0.03) & 366.61(0.00) \\ 
   \hline
\end{tabular}
\end{table}

The good approximations of CVIC by iIS may not be surprising, because our derivation in Section \ref{sec:iIS2} has shown their equivalence in these models. It is surprising to note that the heuristic iWAIC also gives estimates very close to CVIC for model \eqref{eqn:ldmodel1} and \eqref{eqn:ldmodel2}, which contain actually correlated random effects. Furthermore, note that  iWAIC has smaller standard deviations and biases than iIS.   Therefore,  the equivalence of iWAIC to iIS (or CVIC) deserves more empirical and theoretical investigations in the future.

\subsection{CV Posterior p-values in Logistic Regression for Seeds Data} \label{sec:logit}

We consider comparing different methods for computing posterior p-values for identifying outliers in applying logistic regression with random effects to \texttt{Seeds} data, a classic example of WinBUGS (\url{http://www.mrc-bsu.cam.ac.uk/bugs/winbugs/Vol1.pdf}). We obtained the data set from the previous link. The example is taken from Table 3 of \cite{crowder1978beta}. The study concerns about the proportion of seeds that germinated on each of 21 plates arranged according to a 2 by 2 factorial layout by seed and type of root extract.  For $i=1,\ldots, 21$, let $r_i$ be the number of germinated seeds in the $i$th plate, $n_i$ be the total number of seeds in the $i$th plate, $x_{i1}$ be the seed type (0/1), and $x_{i2}$ be root extract (0/1).  The conditional distribution of $r_i$ given $n_i$, $x_{i1}$ and $x_{i2}$ are specified as follows: 
\begin{eqnarray}
r_i | n_i, p_i &\sim& \mbox{Binomial}(n_i, p_i),\\
\mbox{logit}(p_i) &=& \alpha_0 + \alpha_1 x_{1i} + \alpha_2 x_{2i} + \alpha_{12} x_{1i}x_{2i} + b_i,  \\
b_i &\sim&N(0, \sigma^2),
\end{eqnarray}
and parameters $\alpha_0, \alpha_1, \alpha_2, \alpha_{12}$ are assigned with $N(0, 10^6)$ as prior, and $\sigma^2$ is assigned with Inverse-Gamma (0.001, 0.001) as prior. The above model is a member of Bayesian latent variable models depicted by Figure \ref{fig:blvm}. The observable variable is $r_i$, the latent variable is $b_i$,  the covariate variable vector is $(n_i, x_{i1}, x_{i2})$, and the model parameter vector $\btheta$ is $(\alpha_0, \alpha_1, \alpha_2, \alpha_{12})$. We used JAGS to run MCMC for fitting the above model to the \texttt{Seeds} data. For each simulation, we ran 5 parallel chains, each running 1000 iterations for adapting, 2500 iterations for burning in, and 10000 iterations for sampling. 

\begin{figure}[htp]

\caption{Scatterplots of estimated posterior p-values from an MCMC simulation against actual CV posterior p-values. The number for points show indices of plates} \label{fig:ppv}

\centering

\subfloat[Posterior checking]
{\includegraphics[width=0.4\textwidth]{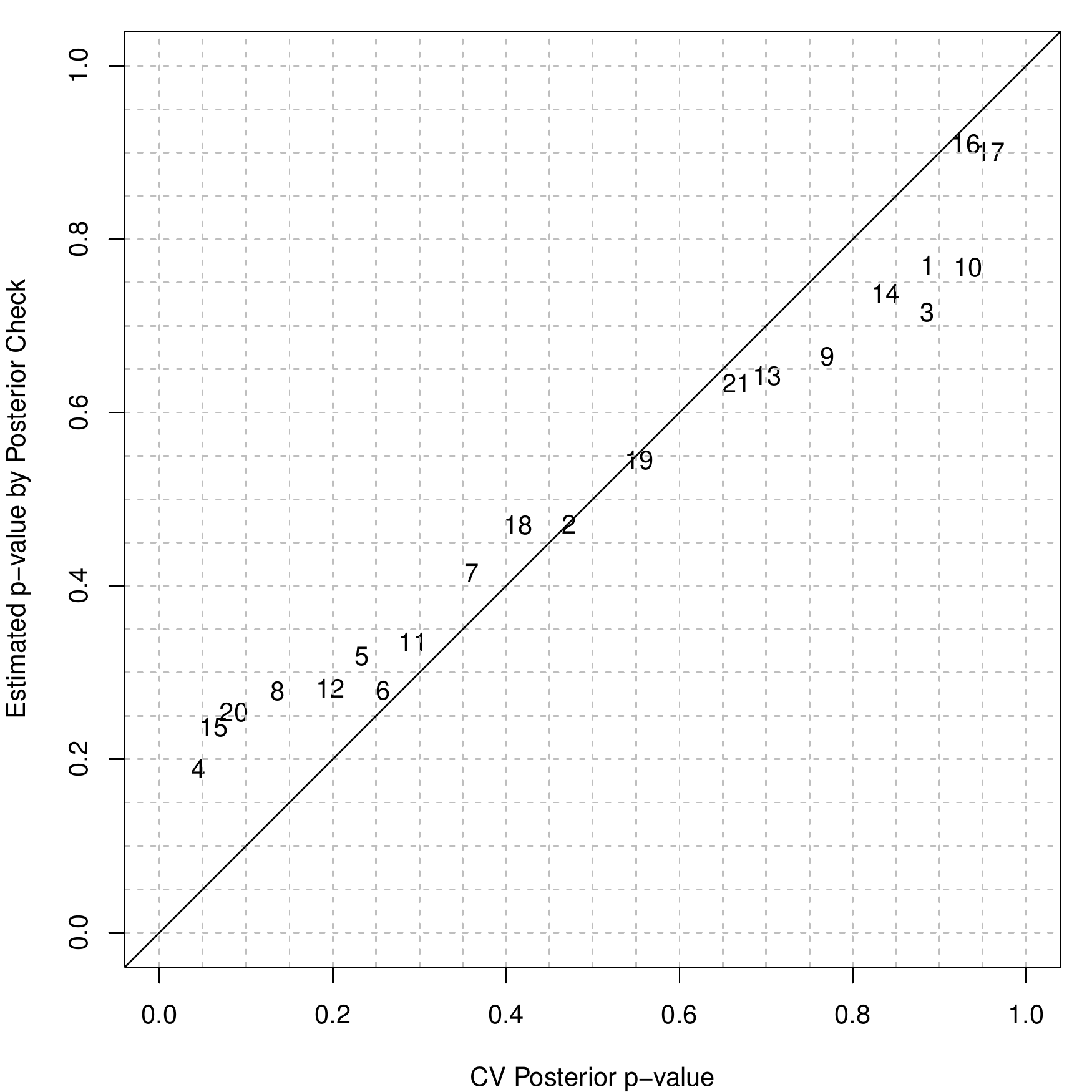}}
\subfloat[Ghosting method]
{\label{fig:ghppv}\includegraphics[width=0.4\textwidth]{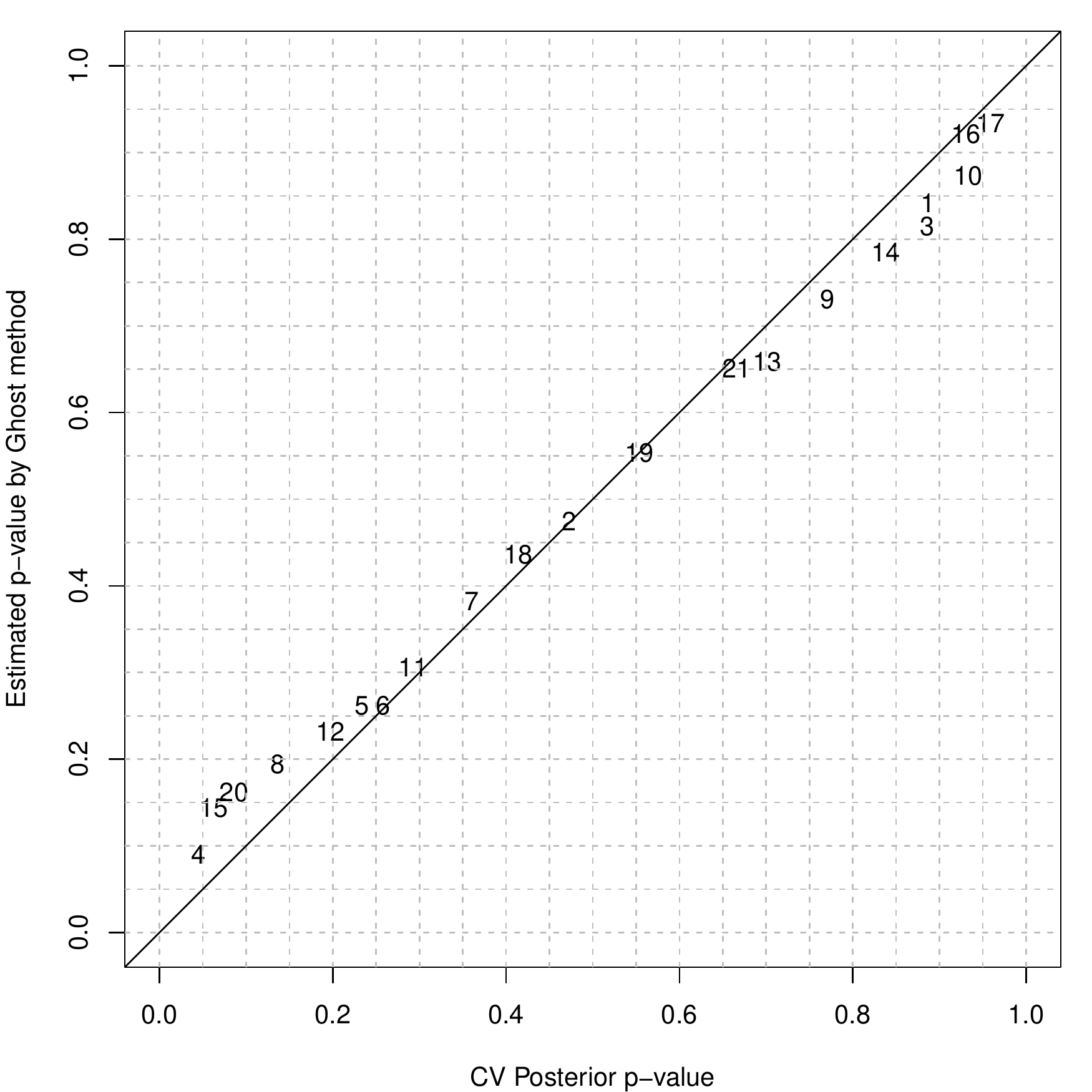}}

\subfloat[Non-integrated IS (nIS)]
{\label{fig:nisppv}\includegraphics[width=0.4\textwidth]{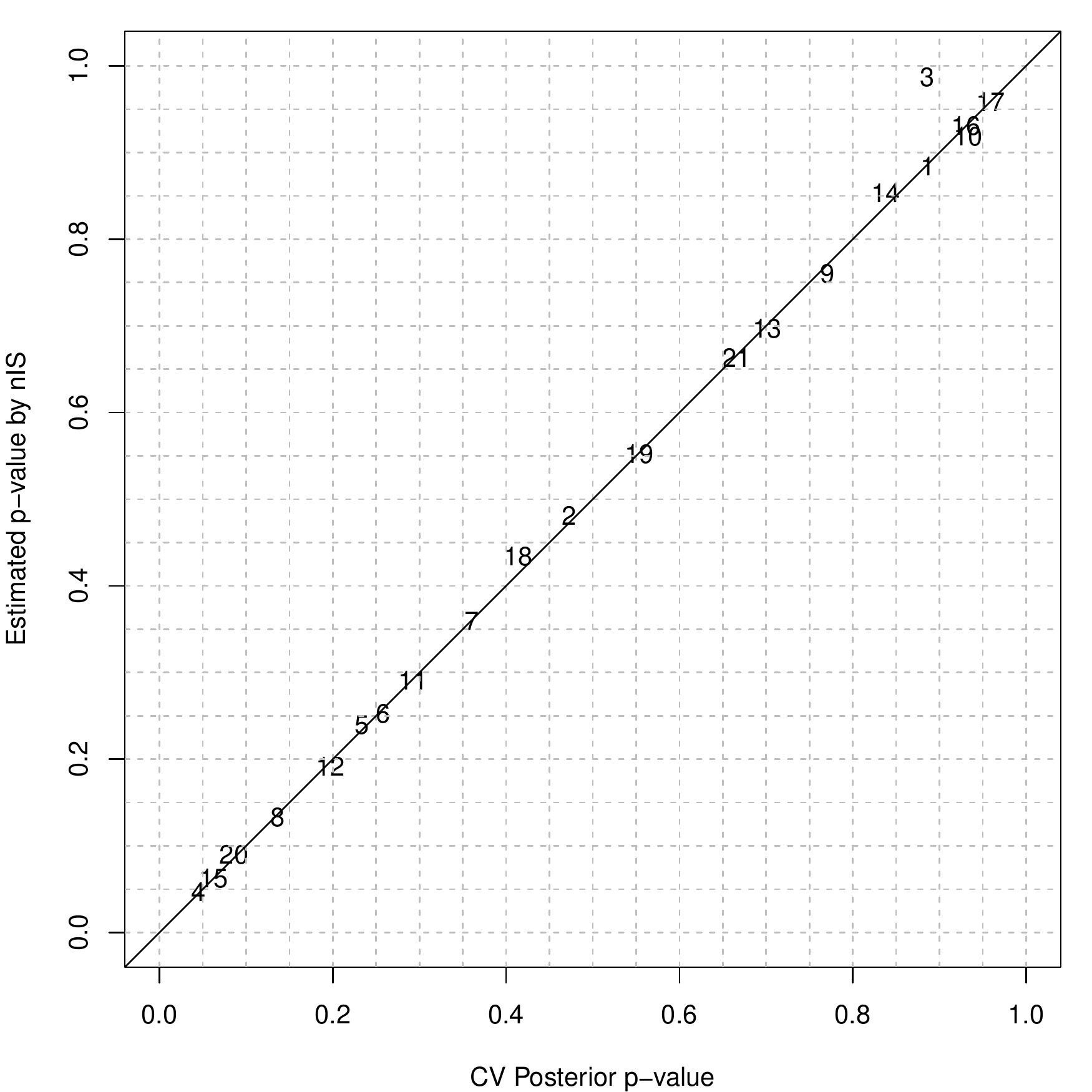}}
\subfloat[Integrated IS (iIS)]
{\includegraphics[width=0.4\textwidth]{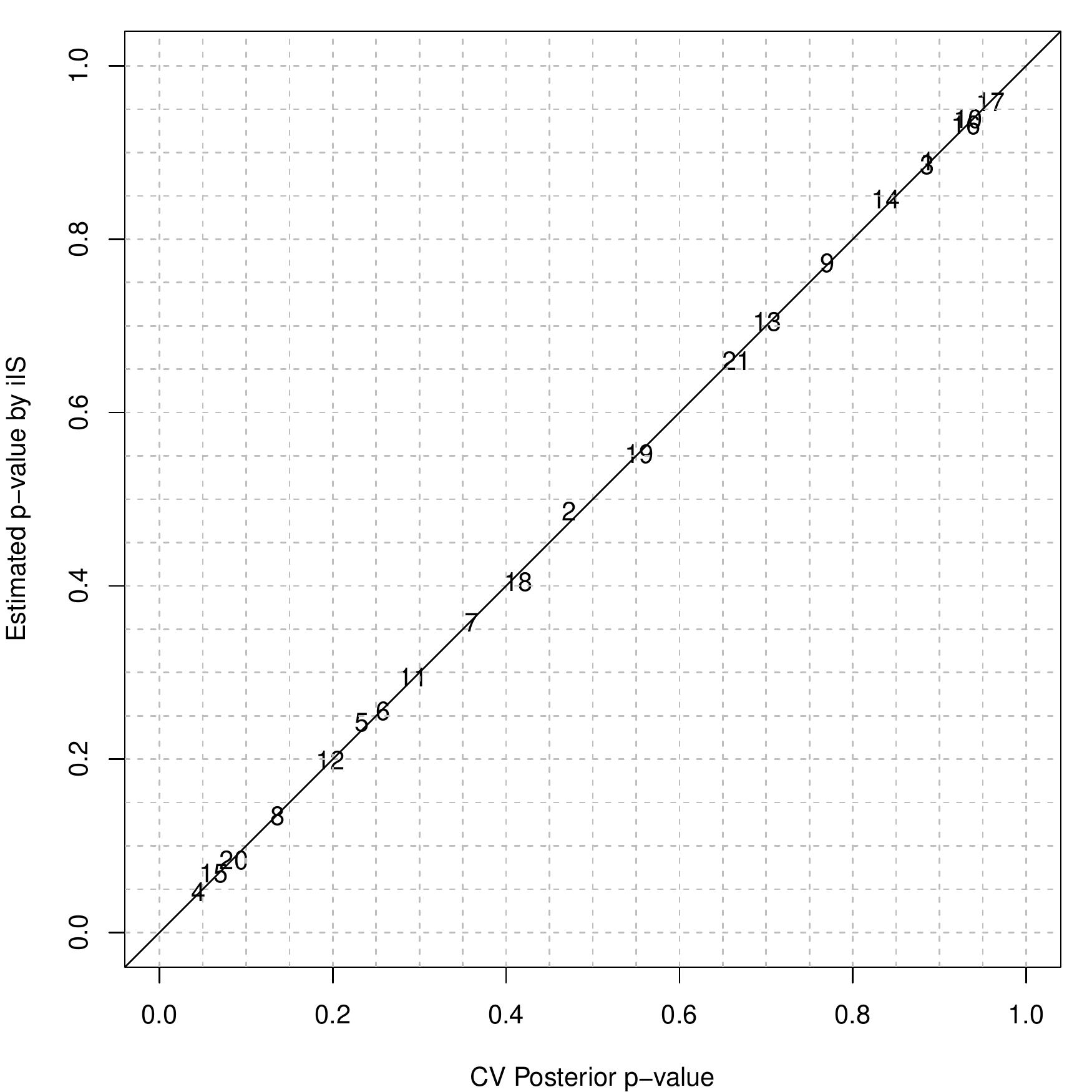}}

\end{figure}

The p-value (given parameters and latent variable) defined by \eqref{eqn:ppvpw} for this example is the right tail probability of Binomial distribution with number of trials $n_i$ and success rate $p_i$:
\begin{equation}
 \mbox{p-value} (r_i\obs, \btheta, b_i) = 1- \mbox{pbinom} (r_i\obs;n_i, p_i) + 0.5\, \mbox{dbinom} (r_i\obs;n_i, p_i),
\end{equation}
where $r_i\obs$ is the actual observation of $r_i$, and pbinom and dbinom denote CDF and PMF of Binomial distribution. Very small or very large p-values indicate that the actual observed $r_i\obs$ falls on the tails of (ie, is unusual to) Binomial ($n_i$, $p_i$).  CV posterior p-value \citep{SIM:SIM1403} for observation $r_i\obs$  is the mean of $\mbox{p-value} (r_i\obs, \btheta, b_i)$ with respect to the CV posterior distribution $P(\btheta, b_i |\mb r\noi\obs)$. If we get a very small or very large CV posterior p-value for observation $r_i\obs$, it indicates that $r_i\obs$ is unusual to the predictive distribution of $r_i$ given $\mb r\noi\obs$.  For this example, when CV posterior p-value for $r_i\obs$ is very small or very large, the germination rate, $r_i\obs$, of the $i$th plate is probably an outlier to other plates. \citet{marshall2007identifying} showed that the CV posterior p-values are uniformly distributed on interval $(0,1)$. We ran actual CV MCMC simulations to find the CV posterior p-values for each of the 21 plates, and the results are displayed by the x-axis in plots of Figure \ref{fig:ppv}. 

We compared four different methods for computing posterior p-values for identifying outliers with only a single MCMC simulation based on the full data set. One method is to apply posterior checking idea of \citet{gelman1996posterior} without considering bias-correction, that is, to average each p-value$(r_i\obs, \btheta, \bb_i)$ with respect to the posterior of $(\btheta, b_i)$ given the full data set $\mb r_{1:21}\obs$.  We will call this method by \textit{posterior checking}. \citet{gelman1996posterior} do not recommend this use of posterior checking because it uses data twice in model building and assessment. However, this method is convenient and so perhaps used very often in practice. Therefore, we include it in comparison. To reduce the bias of including $r_i\obs$ in model fitting,  \citet{SIM:SIM1403} propose \textit{ghosting method}: for each MCMC sample, one averages p-value$(r_i\obs, \btheta, b_i)$ with respect to the conditional distribution of $b_i$ given $\btheta$ (but without $r_i\obs$) to obtain ghosting p-value (which can be done with Monte carlo method by re-generating $b_i$ given $\btheta$ with no reference to $r_i\obs$), then  averages the ghosting p-values over all MCMC samples.  The third method is \textit{non-integrated importance sampling} method (nIS) that  averages p-value$(r_i\obs, \btheta, b_i)$ after being weighted with the inverse of probability density (mass) of $r_i\obs$: $1/\mbox{dbinom}(r_i\obs;n_i, p_i)$. The fourth method is \textit{integrated importance sampling} (iIS). For each MCMC sample, we first average both p-value$(r_i\obs, \btheta, b_i)$ and $\mbox{dbinom}(r_i\obs;n_i, p_i)$ with respect to $P(b_i|\btheta)$ to find the integrated evaluation p-value (equation \eqref{eqn:estA}) and the integrated predictive density (equation \eqref{eqn:inpd}) respectively, then compute the weighted average of the integrated p-values with the reversed integrated predictive density as weights over all MCMC samples using formula \eqref{eqn:iseM}. We can see that the way to obtain ghosting p-value is the same as finding integrated p-value in \eqref{eqn:estA} when $\bb_{1:n}$ are independent given $\btheta$, but without using the reversed integrated density to correct for the optimistic bias in full data posterior of parameters. Therefore,  ghosting method can be viewed as a partial implementation of iIS method presented here.

We calculated 21 posterior p-values with the four method given a MCMC simulation based on the full data set, and repeated this calculation for 100 independent MCMC simulations.  For computing integrated p-values and predictive densities as needed by nIS and ghosting method, we generated 30 of $b_{i}$ from $N(0, \sigma^{2})$ for each plate and each MCMC sample. Figure \ref{fig:ppv} shows the scatter-plots of four sets of estimated posterior p-values given by four different methods against the actual CV posterior p-values from one MCMC simulation. From the figure, we see that the p-values given by posterior checking are more concentrated around 0.5 than the actual CV posterior p-values, and do not appear to be uniformly distributed \citep{gelman2013understanding}. This is because in computing each p-value, the observed value $r_i\obs$ itself is included in model fitting, resulting in optimistic bias. Ghosting method reduces the bias, hence the estimated p-values are closer to the actual CV p-values, and more spread out over $(0,1)$. However, for this example, the bias is still visible from Figure \eqref{fig:ghppv}. Both nIS and iIS give estimates that are very close to the actual values found by CV. However, nIS is less stable than iIS, and sometimes gives very poor estimates; for example the 3rd plate shown in Figure \eqref{fig:nisppv}. 

To measure more precisely the accuracy of estimated p-values to the actual CV p-values, we use absolute relative  error  in percentage scale defined as 
\begin{equation}
\mbox{RE} = (1/n)\sum_{i=1}^{n}\frac{|\hat p_i - p_i|}{\min(p_i,\ 1-p_i)} \times 100,  
\end{equation}
where $\hat {\mb p}_{1:n}$ are estimates of ${\mb p}_{1:n}$. This measure emphasizes greatly on the error between $\hat p_i$ and $p_i$ when $p_i$ is very small or very large, for which we demand more on absolute error than when $p_i$ is close to 0.5. A similar measure (only using $p_i$ in denominator) is used by \citet{marshall2007identifying}. Here we modify the denominator because large p-values are important too. Table \ref{tbl:ppv-er} shows the averages of REs over 100 independent simulations for each method. Clearly, we see that iIS is the best among the four, and improve significantly ghosting and posterior checking methods.  

% !TEX root =  iis.tex
% latex table generated in R 2.15.3 by xtable 1.7-1 package
% Thu Sep  5 11:52:11 2013
\begin{table}[htp]
\caption{Comparisons of the averages of 100 absolute relative errors  (in percentage) of estimated CV p-values from 100 independent MCMC simulations, for logistic regression example. The numbers in brackets indicate standard deviations.} \label{tbl:ppv-er}

\centering
\begin{tabular}{llll}
  \hline
iIS & nIS & Ghosting & Posterior checking \\ 
  \hline
2.319(0.399) & 5.234(1.083) & 35.610(1.267) & 93.887(3.854) \\ 
   \hline
\end{tabular}
\end{table}

\section{Conclusions and Discussions} \label{sec:conc}

In this article, we have introduced two methods (iIS and iWAIC) for approximating leave-one-out cross-validatory predictive evaluations for models with unit-specific and possibly correlated latent variables. The innovation in iIS and iWAIC is that we replace the non-integrated predictive density and evaluation functions by the integrated predictive density and evaluation functions.  iIS is applicable to models with non-identifiable parametrization for which DIC may not be suitable; and also applicable to models with correlated latent variables for which WAIC is not. The extent of applicability of iWAIC remains to be investigated. We have tested iIS and iWAIC in four examples, in which iIS and/or iWAIC provide almost identical approximates to what given by actual leave-one-out cross-validation, whereas other methods show large discrepancies.   In addition, we have found that iWAIC is slightly more stable than iIS.  

Although our empirical results show that iIS and iWAIC provide better approximates of CVIC than DIC, we notice that the implementations of iIS and iWAIC are much more complicated, and requires users to have basic knowledge in statistics and scientific computing (for example a degree in statistics). For the moment, we do not know how to automate their applications as DIC, which can be embedded into a black-box MCMC sampler program. This is a direction for future research one can pursue. 
 
Applicability of iWAIC to models with correlated latent variables still requires more empirical and theoretical investigations. The results of our empirical studies in the lip cancer data give an example that iWAIC provides very close approximates to CVIC.  However, we have to be cautious in the generalization of iWAIC to other models and problems. In the future, we will empirically test iWAIC in many other models using correlated latent variables, for example, the stochastic volatility models used for modelling financial time sequences \citep{jacquier2002bayesian, gander2007stochastic},  multivariate spatial models \citep{feng2012joint}, and many other models considering spatial and temporal correlations \citep{waller1997hierarchical}. We will also investigate iWAIC theoretically, probably using the tools for singular statistical models developed by \citet{watanabe2009algebraic}. 

There is also much research work needed to generalize and extend iIS and iWAIC. We have only shown how to integrate latent variables away in the models where they are unit-specific to improve ordinary nIS and nWAIC.  In many models, a latent variable is shared by many observations. It is still unclear to us how to improve nIS and nWAIC in such models. More ambitiously, we may wonder whether there is a method that requires little technical work but provides very good predictive evaluation for all Bayesian models. 

The insensitivity of CVIC is another important problem that demands further research, as we discuss in Section \ref{sec:simmix}. One may consider other evaluation function than log predictive density for capturing sensitively the difference among models. One may also consider other methods for comparing two sets of log predictive densities resulting from two competing models. However, we think that the method we present in this article for latent variable models may be generally useful for providing better approximation of cross-validatory quantities regardless of the choice of evaluation function. 

\appendix

\section*{Appendices}

\section {Working procedure of iIS}

\begin{enumerate}
 \item Generate MCMC samples $\{(\btheta\mcs, \bb\wi\mcs)$; s= 1,\ldots,S\} from $\PpostfnoM$
 \item For each $s = 1,\ldots,S$
 \begin{enumerate}
 
    \item For each $i = 1,\ldots, n$, generate $\{\bb_i^{(s,r)}; r = 1,\ldots, R\}$ from $P(\bb_i|\bb\noi\mcs,\btheta\mcs)$, and estimate $P(\by_i\obs|\btheta,\bb\noi)$ by
    \begin{equation}
     \label{eqn:estinpd}
     \hat{P} (\by_i\obs|\btheta\mcs, \bb\noi\mcs) = (1/R)\sum_{r=1}^R P(\by_i\obs|\btheta\mcs,\bb\noi\mcs, \bb_i^{(s,r)}).
    \end{equation}
    Then, we can find iIS weight:
    \begin{equation}
    \label{eqn:estW}
    W_{i}\iIS (\btheta\mcs,\bb\noi\mcs) = \dfrac{1}{\hat{P} (\by_i\obs|\btheta\mcs, \bb\noi\mcs)}. 
    \end{equation}
    \item For each $i = 1,\ldots, n$, generate $\{\tilde\bb_i^{(s,k)}; k = 1,\ldots, K\}$ from $P(\bb_i|\bb\noi\mcs,\btheta\mcs)$, and estimate integrated evaluation function $A$ by
    \begin{equation}
    \label{eqn:estA}
    A (\by_i\obs,\btheta\mcs,\bb\noi\mcs) = (1/K)\sum_{k=1}^K a\left(\by_i\obs,\btheta\mcs,\tilde\bb_i^{(s,k)}\right).  
    \end{equation}

 \end{enumerate}

    \item Estimate expected evaluation function $a$ with respect to $P(\btheta,\bb_{1:n}|\by\noi\obs)$ by
    \begin{equation}
    \label{eqn:estEA}
      \hat {E} \iIS \cvpostnoM (a(\by_i\obs,\btheta,\bb_i)) = \dfrac{(1/S)\sum_{s=1}^S \left[A (\by_i\obs,\btheta\mcs,\bb\noi\mcs)W_{i}\iIS (\btheta\mcs,\bb\noi\mcs)\right]}{(1/S)\sum_{s=1}^S W_{i}\iIS (\btheta\mcs,\bb\noi\mcs)}.
    \end{equation}

\end{enumerate}

Note that, if we are only interested in computing CVIC, don't need to do step 2(b), and take the numerator in \eqref{eqn:estEA} to be 1 as warranted by theory.

\section {Working procedure of iWAIC}

\begin{enumerate}
 \item Generate MCMC samples $\{(\btheta\mcs, \bb\wi\mcs)$; s= 1,\ldots,S\} from $\PpostfnoM$
 \item For each $s = 1,\ldots,S$ and each $i = 1,\ldots, n$, generate $\{\bb_i^{(s,r)}; r = 1,\ldots, R\}$ from $P(\bb_i|\bb\noi\mcs,\btheta\mcs)$, and estimate integrated predictive density $P(\by_i\obs|\btheta,\bb\noi)$ by
    \begin{equation}
     \hat{P} (\by_i\obs|\btheta\mcs, \bb\noi\mcs) = (1/R)\sum_{r=1}^R P(\by_i\obs|\btheta\mcs,\bb\noi\mcs, \bb_i^{(s,r)}).
    \end{equation}

\item Estimate log CV posterior predictive density: 
    \begin{equation}
    \log (\hat P(\by_i\obs|\by\noi\obs)) = \log((1/S)\sum_{s=1}^S \hat{P} (\by_i\obs|\btheta\mcs, \bb\noi\mcs)) - V_{s=1}^S \log(\hat{P} (\by_i\obs|\btheta\mcs, \bb\noi\mcs)),
    \end{equation}
where $V_{s=1}^S a\mcs$ stands for sample variance of $(a^{(1)},\ldots, a^{(S)}$).

\item Find iWAIC:
\begin{equation}
\mbox{iWAIC} = -2\sum_{i=1}^n \log (\hat P(\by_i\obs|\by\noi\obs)).
\end{equation}

\end{enumerate}

\onehalfspacing

\bibliographystyle{asa}

\bibliography{zlibrary,library}

\end{document}